\documentclass{article}
\usepackage{hyperref}
\usepackage{amsmath}
\usepackage{graphicx}
\usepackage{amsfonts}
\usepackage{amssymb}
\renewcommand{\thefootnote}{\arabic{footnote}}

\begin{document}

\title{An Embedding of the BV Quantization into an N=1 Local Superfield Formalism}
\author{\textsc{D.M.~Gitman}$^{a)}$\thanks{E-mail: gitman@dfn.if.usp.br},
\textsc{P.Yu.Moshin}$^{a),b)}$\thanks{E-mail: moshin@dfn.if.usp.br}, and
\textsc{A.A. Reshetnyak}$^{b)}$\thanks{E-mail: reshet@tspu.edu.ru}\\ \\$^{a)}$Instituto
de F\'{\i}sica, Universidade de S\~{a}o Paulo,\\Caixa Postal 66318-CEP, 05315-970 S\~{a}o Paulo, S.P.,
Brazil\\$^{b)}$Tomsk State Pedagogical University, 634041 Tomsk, Russia}
\date{}
\maketitle

\begin{abstract}
We propose an $N=1$ superfield formulation of Lagrangian quantization in
general hypergauges by extending a reducible gauge theory to a superfield
model with a local dependence on a Grassmann parameter $\theta$.
By means of $\theta$-lo\-cal functions of the quantum and ga\-uge-fi\-xing
actions in terms of Darboux coordinates on the antisymplectic
manifold, we construct superfield generating functionals of
Green's functions, including the effective action.
We prove the gauge-independence of the S-matrix, obtain
the Ward identities and establish a relation of the proposed
local quantization with the BV method and the multilevel
Batalin--Tyutin formalism.
\end{abstract}

\section{Introduction}

The quantization of gauge theories on the basis of BRST symmetry
\cite{BRST} is usually carried out in the Hamiltonian \cite{BFV}
or Lagrangian \cite{BV} schemes, which were recently given a
superfield description \cite{BatalinBeringDamgaard1,LMR} based on
nontrivial \cite{BatalinBeringDamgaard1} and trivial \cite{LMR}
relations between the even $t$ and odd $\theta$ components of
supertime. These works realize a geometric interpretation of BRST
transformations in terms of supertranslations, which originally
provided a basis for the superspace formulation
\cite{BonoraToninBaulieu} of quantum theories of Yang--Mills type
\cite{Hull}, and, in a larger context, were applied to a classical
and quantum description
\cite{AlexandrovKontsevichSchwarzZaboronsky,BatalinMarnelius1,Gozzi}
of generalized Poisson sigma-models \cite{CattaneoFelder} and
$D=1$ sigma-models with an arbitrary $N\geq1$ number of Grassmann
coordinates, as well as to a construction \cite{Gozzi} of the
partition function with $N=2$ (for more details, see
\cite{BatalinMarnelius2}).

The Lagrangian formalism \cite{LMR} is a superfield modification
of the BV method including non-Abelian hypergauges
\cite{BatalinTyutin}. The formalism \cite{LMR} provides a
relatively complete insight into superfield quantization
based on the properties of solutions to the
generating equations;
however, it does not indicate a detailed relation between these solutions and a gauge
theory. It is therefore natural to complement the
formalism by an \emph{explicit superfield description} of the
gauge algebra for a given model. This
problem has so far remained open. Thus, the definition of a
classical action of superfields,
${\mathcal{A}}^{i}(\theta)=A^{i}+{\lambda}^{i}\theta$, on a
superspace with coordinates $(x^{\mu},\theta)$,
$\mu=0,\ldots,D-1$, as an integral of a nontrivial\footnote{A
trivial density ${\mathcal{L}}(x,\theta)$
is understood in the form $\int d^{D}x\,d\theta\,{\mathcal{L}%
}(x,\theta)=\int d\theta\,\theta\,S_{0}\left(  \mathcal{A}(\theta)\right)
=S_{0}(A),$ where $S_{0}(A)$\ is a usual classical action.}
odd density
${\mathcal{L}}(x,\theta)$ is a question for every given model. As a
consequence, the vacuum functional $Z$ and generating functional
of Green's functions $Z[\Phi^{\ast}]$ of \cite{LMR} exhibit a
peculiarity. Namely, these objects differ from their counterparts
of the BV \cite{BV} and Batalin--Tyutin
\cite{BatalinTyutin} methods, which is implied by a
dependence of the gauge
fermion $\Psi\lbrack\Phi]$ and quantum action ${S}[\Phi,{\Phi}%
^{\ast}]$ on the components $\lambda^{A}$ of superfields
$\Phi^{A}(\theta)$ in the multiplet
$(\Phi^{A},{\Phi}_{A}^{\ast
})(\theta)=(\phi^{A}+\lambda^{A}\theta,\phi_{A}^{\ast}-\theta
J_{A})$, where $(\phi^{A},\phi_{A}^{\ast},\lambda^{A},J_{A})$
constitute the complete set of variables in the BV formalism.

The aim of this paper is to propose an $N=1$ local superfield
Lagrangian quantization in which the quantities of
an initial classical theory are realized through a
$\theta$\emph{-local superfield model} (LSM). Note that we adopt
a terminology consistent with that of
the papers \cite{BatalinBeringDamgaard2}, in which the quantization
\cite{BatalinBeringDamgaard1} with a single fermion supercharge
$Q(t,\theta )$ containing the BRST charge and unitarizing
Hamiltonian was extended to $N=2$ (non-spacetime)
supersymmetries, and then to an arbitrary number of
supercharges, $Q^{k}(t,\theta^{1},...,\theta^{N})$, $k=1,...,N$,
depending on Grassmann variables $\theta^{k}$. An LSM presents
the objects of a gauge theory
in terms of $\theta$-local functions trivially (in comparison with
the Hamiltonian superfield scheme
\cite{BatalinBeringDamgaard1}) related to the spacetime
coordinates, which means that the derivatives over the
even $t$ and odd $\theta$ components of supertime are taken
independently. This implies an analogy with classical field theory
in which, as distinct from the tradition inspired by a superfield
description of the SUSY spacetime (used in
\cite{BatalinBeringDamgaard1,LMR}
to define the classical and quantum actions as functionals in superspace),
we start from $\theta $-local functions.
Our arguments are based on the fact that the $t$-local quantities
$\Gamma^{\mathbf{p}}(t)$, $H(t)$, $t$ and the operation
$\{\,\cdot\,,\,\cdot\,\}$, being the phase-space coordinates,
Hamilton function, even time and Poisson bracket of a $t$-local
field theory, are put into a correspondence with the quantities
$\Gamma^{p}$, $\{\mathbf{p},t\}\subset p$, $S(\Gamma)$, $S\supset
H(t)$, $\mu$ and the operation $(\,\cdot\,,\,\cdot\,)$, being the
field-antifield coordinates, quantum action, odd parameter of BRST
transformations and antibracket used in the BV method. Extending
this duality, we consider (replacing $\mu$ by a parameter $\theta$
whose range includes $\mu$) the latter quantities and operation
(as well as the generating functionals of Green's functions) as
$\theta$-local objects, so that, in view of the nilpotency of
$\theta$, physical quantities are determined only by the
$\theta$-independent part of $\theta$-local
functions. Using these principles, we reproduce the dynamics
and gauge invariance
of an initial theory (with $\theta=0$) in terms of $\theta$-local
equations, called \emph{Lagrangian }and
\emph{ Hamiltonian }systems (LS, HS) with a \emph{dynamical} odd
time $\theta $, implying that $\theta$ generally enters an LS or
HS through a differential operator $\partial_{\theta}$ which
describes the $\theta $-evolution.
The proposed formalism permits us to circumvent the mentioned peculiarity
of the functionals $Z$, $Z[\Phi^{\ast}]$ in \cite{LMR} and to solve
the following problems:
We develop a \emph{dual description} that interrelates the
Lagrangian \cite{BV} and Hamiltonian \cite{BFV} formulations of an
arbitrary reducible LSM proposed in \cite{GrigorievDamgaard}
for irreducible gauge theories (with bosonic classical fields and
gauge parameters) in terms of a BRST charge
related to a formal dynamical system with first-class
constraints of a higher stage of reducibility. An HS constructed
from $\theta$-local quantities (a quantum action, a
gauge-fixing action, and an arbitrary bosonic function) encodes,
through a $\theta$-local antibracket, both BRST and
anticanonical-like transformations in terms of a universal set of
equations underlying the gauge-independence of the S-matrix.
For the first time within superfield quantization, we introduce a
\emph{superfield effective action} (also in the case of
non-Abelian hypergauges). We establish a relation of the proposed
local quantization with the BV and Batalin--Tyutin methods
\cite{BV,BatalinTyutin}, as well as with the superfield formalism
\cite{LMR}.

We use DeWitt's condensed notation and the conventions of \cite{LMR}.
As usual, the rank of an even $\theta$-local supermatrix $M(\theta)$
with $Z_{2}$-grading $\varepsilon$ is characterized by a pair of
numbers $\overline {m}=(m_{+},m_{-})$, where $m_{+}$ ($m_{-}$) is
the rank of the Bose--Bose (Fermi--Fermi) block of the
$\theta$-independent part of the supermatrix $M(\theta)$,
$\mathrm{rank}\Vert M(\theta)\Vert=\mathrm{rank}\Vert M(0)\Vert$.
With respect to the same Grassmann parity $\varepsilon$, we
understand the dimension of a smooth supersurface, also
characterized by a pair of numbers in the sense of the definition
\cite{Berezin} of a supermanifold, so that the pair
$(m_{+},m_{-})$ coincides with the corresponding numbers of the
Bose and Fermi components of $z^{i}(0)$, being the $\theta$-independent
parts of local coordinates $z^{i}(\theta)$ parameterizing this
supersurface.

\section{Classical Description of a $\theta$-local Superfield Model}

In this section, we propose odd Lagrangian and
Hamiltonian descriptions of an LSM as extensions of a
usual model of classical fields $A^{i}$,
$i=1,...,n=n_{+}+n_{-}$, to $\theta$-local theories defined
on the respective tangent $\Pi T\mathcal{M}_{\mathrm{CL}}$ =
$\left\{ \mathcal{A}^{I},\partial_{\theta}\mathcal{A}^{I}\right\}
(\theta)$ and cotangent  $\Pi T^{\ast}\mathcal{M}_{\mathrm{CL}}$
= $\left\{\Gamma^P_{\mathrm{CL}}= \left(\mathcal{A}^{I},
\mathcal{A}^{\ast}_{I}\right)\right\} (\theta)$,
$I=1,\ldots,N=N_{+}+N_{-}$ odd bundles,\footnote{$\Pi$ denotes the operation
that changes the coordinates of a (co)tangent fiber bundle
$T^{(\ast)}\mathcal{M}_{\mathrm{CL}}$ over a configuration
$\mathcal{A}^{I}$ into the coordinates of the opposite Grassmann
parity, whereas $N_{+}$, $N_{-}$ are the numbers of bosonic and
fermionic fields, among which there may be superfields
corresponding to the ghosts of the minimal sector in the BV
quantization scheme.} $(n_{+},n_{-})\leq
(N_{+},N_{-})$. The superfields $(\mathcal{A}^{I},\partial_{\theta}%
\mathcal{A}^{I})(\theta)$ and superantifields
$\mathcal{A}_{I}^{\ast }(\theta)$, $\mathcal{A}_{I}^{\ast
}(\theta)=(A_{I}^{\ast}-\theta J_{I})$, are defined in a superspace $\mathcal{M}%
=\widetilde{\mathcal{M}}\times\widetilde{P}$ parameterized by
$\left( z^{M},\theta\right)$, where the  coordinates
$z^{M}\subset i\subset I$ include Lorentz vectors and spinors of
the superspace $\widetilde{\mathcal{M}}$.
The basic objects of the odd Lagrangian and Hamiltonian formulations
of an LSM are \emph{Lagrangian} and \emph{Hamiltonian} actions, $\left[S_{\mathrm{L}},%
S_{\mathrm{H}}\right]$: $\left[\Pi T\mathcal{M}%
_{\mathrm{CL}}\times\{\theta\}, \Pi T^{\ast
}\mathcal{M}_{\mathrm{CL}}\times\{\theta\}\right]$ $\rightarrow$ $\Lambda_{1}(\theta;\mathbb{R})$,
being a respective $C^{\infty}(\Pi
T\mathcal{M}_{\mathrm{CL}})$- and $C^{\infty}(\Pi
T^{\ast}\mathcal{M}_{\mathrm{CL}})$-functions taking values in a
real Grassmann algebra $\Lambda_{1}(\theta;\mathbb{R})$. The
actions determine the respective functionals $Z_{\mathrm{L}}[\mathcal{A}]$ and
$Z_{\mathrm{H}}[{\Gamma}_{k}]$, whose $\theta$-densities are defined with
accuracy up to arbitrary functions
$\left[g((\mathcal{A},\partial_{\theta
}\mathcal{A})(\theta),\theta), g_{\mathrm{H}}((\mathcal{A},
\mathcal{A}^{\ast})(\theta),\theta)\right] \in
{\ker}\{\partial_{\theta}\}$, $\vec {\varepsilon}(g) = \vec
{\varepsilon}(g_{\mathrm{H}}) =\vec{0}$ $(\int d\theta = \partial^l_{\theta} \equiv \partial_{\theta})$%
\begin{eqnarray}
&& Z_{\mathrm{L}}[\mathcal{A}]=\partial_{\theta}S_{\mathrm{L}}(\theta),\;
Z_{\mathrm{H}}[{\Gamma}_{k}]= \partial_{\theta}\left[
V_{P}^{k}(\Gamma
(\theta))\partial_{\theta}^{r}{\Gamma}_{k}^{P}(\theta)-S_{\mathrm{H}}%
(\Gamma_{k}(\theta),\theta)\right], k=\mathrm{CL} \nonumber \\
&& \vec{\varepsilon }(Z_{\mathrm{L}}) = \vec{\varepsilon
}(Z_{\mathrm{H}}) =\vec{\varepsilon}(\theta)=(1,0,1),\ \vec{\varepsilon}(S_{\mathrm{L}%
})=\vec{\varepsilon}(S_{\mathrm{H}%
})=\vec{0}, \label{1}%
\end{eqnarray}
where $Z_{\mathrm{H}}[{\Gamma}_{k}]$ is expressed in terms of  an
antisymplectic potential,
$V_{P}^{k}(\Gamma(\theta))=1/2(\Gamma^{Q}\omega_{QP}^{k})(\theta)$,
related to a flat antisymplectic metric
$\omega_{PQ}^{k}(\theta)$ and an odd Poisson bivector
$\omega_{k}^{PQ}(\theta)$:
$\omega_{k}^{PD}(\theta)\omega_{DQ}^{k}(\theta)=\delta^{P}{}_{Q}$,
$\omega
_{k}^{PQ}(\theta)\equiv\left(  \Gamma_{k}^{P}(\theta),\Gamma_{k}^{Q}%
(\theta)\right)  _{\theta}$.
The values $\vec{\varepsilon}=(\varepsilon_{P},\varepsilon_{\bar{J}%
},\varepsilon)$, $\varepsilon=\varepsilon_{P}+\varepsilon_{\bar{J}}$, of
a $Z_{2}$-grading introduced in \cite{Reshet1}, with the \emph{auxiliary}
components $\varepsilon_{\bar{J}}$,
$\varepsilon_{P}$ related to the respective coordinates $\left(  z^{M}%
,\theta\right)  $ of the superspace $\mathcal{M}$, are defined by
$\vec{\varepsilon}(\mathcal{A}^{I})=\left( (\varepsilon
_{P})_{I},(\varepsilon_{\bar{J}})_{I},\varepsilon_{I}\right) =
\vec{\varepsilon}(\mathcal{A}^{\ast}_{I}) + (1,0,1)$. Note that
$\mathcal{M}$ can be realized as the quotient of a symmetry
supergroup ${J}=\bar{J}\times P$, $P=\exp(i\mu p_{\theta})$ for
the functional $Z_{\mathrm{L}}[\mathcal{A}]$, where $\mu$ and $p_{\theta}$ are
a nilpotent parameter and a generator of
$\theta$-translations, whereas $\bar{J}$ is chosen as a spacetime
SUSY group. The quantities $\varepsilon_{\bar{J}}$,
$\varepsilon_{P}$ are the Grassmann parities of
coordinates in some representation spaces of $\bar{J}$, $P$.
These objects are introduced for a correct spin-statistic relation
in operator quantization.

Due\thinspace to\thinspace a\thinspace $J$-scalar\thinspace nature\thinspace of\thinspace
$Z_{\mathrm{L}}[\mathcal{A}]$, $Z_{\mathrm{H}}[%
{\Gamma}_{k}]$ it is only $[S_{\mathrm{L}%
},S_{\mathrm{H}%
}](\theta)$, among
$[S_{\mathrm{L}},S_{\mathrm{H}}](\theta),Z_{\mathrm{L}}[\mathcal{A}],Z_{\mathrm{H}}[{\Gamma}_{k}]$,
invariant
under a $J$-superfield representation $T$ restricted
to $\bar{J}$, $\left. T\right| _{\bar{J}}$, that transform
nontrivially
with respect to the total representation $T$ under $\mathcal{A}^{I}%
(\theta)\rightarrow\mathcal{A}^{\prime}{}^{I}(\theta)=(\left.  T\right|
_{\bar{J}}\mathcal{A})^{I}(\theta-\mu)$, for instance,%
\begin{equation}
\delta S_{\mathrm{L}}(\theta)=S_{\mathrm{L}}\left(  \mathcal{A}^{\prime
}(\theta),\partial_{\theta}\mathcal{A}^{\prime}(\theta),{\theta}\right)
-S_{\mathrm{L}}(\theta)=-\mu\left[  {\partial}/{\partial\theta}+P_{0}%
(\theta)(\partial_{\theta}U)(\theta)\right]  S_{\mathrm{L}}(\theta).\label{2}%
\end{equation}
Here, we have introduced the nilpotent operator $(\partial_{\theta}{U}%
)(\theta)=\partial_{\theta}\mathcal{A}^{I}(\theta)\partial_{l}/\partial
\mathcal{A}^{I}(\theta)=[\partial_{\theta},U(\theta)]_{-}$, $U(\theta
)=P_{1}(\theta)\mathcal{A}^{I}(\theta)\partial_{l}/\partial\mathcal{A}%
^{I}(\theta)$, and a set of projectors onto $C^{\infty}(\Pi T^{(\ast)}%
\mathcal{M}_{\mathrm{CL}})$ $\times$ $\{\theta\}$, $\{P_{a}(\theta
)=\delta_{a0}(1-\theta\partial_{\theta})+\delta_{a1}\theta\partial_{\theta}%
$, $a=0,1\}$.

Assuming the existence of critical configurations for $Z_{\mathrm{H}}[{\Gamma}_{k}]$,
$Z_{\mathrm{L}}[\mathcal{A}]$, we present the HS dynamics through a $\theta$-local
antibracket, and the LS dynamics in terms of superfield Euler--Lagrange equations,
\begin{eqnarray}
&& \partial_{\theta}^{r}\Gamma_{\mathrm{CL}}^{P}(\theta)=\left(
\Gamma_{\mathrm{CL}}^{P}(\theta),S_{\mathrm{H}}(\theta)\right)
_{\theta }; \label{3}\\
&&
\frac{\delta_{l}Z_{\mathrm{L}}[\mathcal{A}]}{\delta\mathcal{A}^{I}(\theta)}=\left[
\frac{\partial_{l}}{\partial\mathcal{A}^{I}(\theta)}-(-1)^{\varepsilon_{I}%
}\partial_{\theta}\frac{\partial_{l}}{\partial(\partial_{\theta}\mathcal{A}%
{}^{I}(\theta))}\right]  S_{\mathrm{L}}(\theta)\equiv\mathcal{L}_{I}%
^{l}(\theta)S_{\mathrm{L}}(\theta)=0, \label{4}%
\end{eqnarray}
where the latter system is equivalent (since
$\partial_{\theta}^{2}\mathcal{A}^{I}(\theta)\equiv0$) to an LS
characterized by $2N$ formally second-order differential equations
in $\theta$:%
\vspace{-2ex}
\begin{align}
&  \partial_{\theta}^{2}\mathcal{A}^{J}(\theta)\frac{\partial_{l}%
^{2}S_{\mathrm{L}}(\theta)}{\partial(\partial_{\theta}\mathcal{A}^{I}%
(\theta))\partial(\partial_{\theta}\mathcal{A}^{J}(\theta))}\equiv
\partial_{\theta}^{2}\mathcal{A}^{J}(\theta)(S_{\mathrm{L}}^{\prime\prime
})_{IJ}(\theta)=0,\nonumber\\
&  {\Theta}_{I}(\theta)\equiv\frac{\partial_{l}S_{\mathrm{L}}(\theta
)}{\partial\mathcal{A}^{I}(\theta)}-(-1)^{\varepsilon_{I}}\left[
\frac{\partial}{\partial\theta}\frac{\partial_{l}S_{\mathrm{L}}(\theta
)}{\partial(\partial_{\theta}\mathcal{A}^{I}(\theta))}+(\partial_{\theta
}U)(\theta)\frac{\partial_{l}S_{\mathrm{L}}(\theta)}{\partial(\partial
_{\theta}\mathcal{A}^{I}(\theta))}\right]  =0. \label{5}%
\end{align}
An equivalence of the two descriptions is implied by the
nondegeneracy of the supermatrix $\left\|
(S_{\mathrm{L}}^{\prime\prime})_{IJ}(\theta)\right\|  $ in
(\ref{5}), under a Legendre transformation of
$S_{\mathrm{L}}(\theta)$
with respect to $\partial_{\theta}^{r}\mathcal{A}^{I}(\theta)$,%
\begin{equation}
S_{\mathrm{H}}(\Gamma_{\mathrm{CL}}(\theta),\theta)=\mathcal{A}_{I}^{\ast
}(\theta)\partial_{\theta}^{r}\mathcal{A}^{I}(\theta)-S_{\mathrm{L}}%
(\theta),\;\mathcal{A}_{I}^{\ast}(\theta)={\partial }/{\partial(\partial_{\theta}^{r}\mathcal{A}^{I}(\theta))}
S_{\mathrm{L}}%
(\theta). \label{6}%
\end{equation}
In this case, the equivalence of an LS and HS is guaranteed by
the respective settings ($\theta=0$, $k=\mathrm{CL}$) of
the Cauchy problem for integral curves $\hat{\mathcal{A}}%
^{I}(\theta)$ and $\hat{\Gamma}{}_{k}^{P}(\theta)$, modulo
the continuous part of $I$,
\begin{equation}
\left(  \hat{\mathcal{A}}^{I},\partial_{\theta}^{r}\hat{\mathcal{A}}%
^{I}\right)  (0)=\left(
\overline{\mathcal{A}}{}^{I},\overline{\partial
_{\theta}^{r}\mathcal{A}}{}^{I}\right)
,\;\hat{\Gamma}_{k}^{P}(0)=\left(
\overline{\mathcal{A}}{}^{I},\overline{\mathcal{A}}{}_{I}^{\ast}\right)
:\;\overline{\mathcal{A}}{}_{I}^{\ast}=P_{0}\left[  \frac{\partial
S_{\mathrm{L}}(\theta)}{\partial(\partial_{\theta}^{r}\mathcal{A}^{I}%
(\theta))}\right]  \left(  \overline{\mathcal{A}}{}^{I},\overline
{\partial_{\theta}^{r}\mathcal{A}}{}^{I}\right).\label{7}%
\end{equation}
The \emph{Lagrangian
constraints}{\thinspace}${\Theta}_{I}(\theta)={\Theta
}_{I}(\mathcal{A}(\theta),\partial_{\theta}\mathcal{A}(\theta),\theta
)$  identical to half the HS equations, ${\Theta}_{I}(\theta) =
-(\partial _{\theta}^{r}\mathcal{A}_{I}^{\ast}(\theta)+
S_{\mathrm{H}},_{I}(\theta ))(-1)^{\varepsilon_{I}}$, in view of
transformations (\ref{6}), {\thinspace}restrict{\thinspace}the
setting of the{\thinspace}Cauchy problem for an LS and HS, and may
be functionally dependent as first-order equations in $\theta$.

On condition that there exists (at least locally) a supersurface
$\Sigma\subset\mathcal{M}_{\mathrm{CL}}$ such that%
\begin{equation}
\left.  {\Theta}_{I}(\theta)\right|  _{\Sigma}=0,\;\dim\Sigma=\overline
{M},\;\mathrm{rank}\left\|  \mathcal{L}_{J}^{l}(\theta_{1})\left[
\mathcal{L}_{I}^{l}(\theta_{1})S_{\mathrm{L}}(\theta_{1})(-1)^{\varepsilon
_{I}}\right]  \right\|  _{\Sigma}=\overline{N}-\overline{M},\label{8}%
\end{equation}
there exist $M=M_{+}+M_{-}$ independent identities:%
\begin{equation}
\partial_{\theta}\left[\frac{\delta Z_{\mathrm{L}}[\mathcal{A}]}{\delta\mathcal{A}^{I}(\theta)}%
{\hat{\mathcal{R}}}_{\mathcal{A}_{0}}^{I}(\theta;{\theta}_{0})\right]=0,\;{\hat
{\mathcal{R}}}_{\mathcal{A}_{0}}^{I}(\theta;{\theta}_{0})=\sum\nolimits_{k\geq0}\left(
\left(  \partial_{\theta}\right)
^{k}\delta(\theta-\theta_{0})\right)
{\hat{\mathcal{R}}}_{k}{}_{\mathcal{A}_{0}}^{I}\left(  \mathcal{A}%
(\theta),\partial_{\theta}\mathcal{A}(\theta),\theta\right)  .\label{9}%
\end{equation}
The generators ${\hat{\mathcal{R}}}_{\mathcal{A}_{0}}^{I}(\theta;{\theta}%
_{0})$ of \emph{general gauge transformations},
\[
\delta_{g}\mathcal{A}^{I}(\theta)=\partial_{\theta_{0}}\left[{\hat{\mathcal{R}}%
}_{\mathcal{A}_{0}}^{I}(\theta;{\theta}_{0}){\xi}^{\mathcal{A}_{0}}(\theta
_{0})\right],\;\vec{\varepsilon}({\xi}^{\mathcal{A}_{0}})=\vec{\varepsilon
}_{\mathcal{A}_{0}},\;\mathcal{A}_{0}=1,...,\;M_{0}=M_{0+}+M_{0-},
\]
that leave $Z_{\mathrm{L}}[\mathcal{A}]$ invariant are functionally dependent on the
assumption of locality and $\bar{J}$-covariance, provided that $\mathrm{rank}%
\left\|  \sum_{k\geq0}{\hat{\mathcal{R}}}_{k}{}_{\mathcal{A}_{0}}^{I}%
(\theta)\left(  \partial_{\theta}\right)  ^{k}\right\|  _{\Sigma}=\overline
{M}<\overline{M}_{0}$. The dependence of ${\hat{\mathcal{R}}}_{\mathcal{A}%
_{0}}^{I}(\theta;{\theta}_{0})$ implies the existence (on solutions of an LS)
of proper zero-eigenvalue eigenvectors, ${\hat{\mathcal{Z}}}{}_{\mathcal{A}%
_{1}}^{\mathcal{A}_{0}}\left(  \mathcal{A}(\theta_{0}),\partial_{\theta_{0}%
}\mathcal{A}(\theta_{0}),\theta_{0};\theta_{1}\right)  $, with a
structure analogous to
${\hat{\mathcal{R}}}_{\mathcal{A}_{0}}^{I}(\theta;{\theta}_{0})$
in (\ref{9}), which exhaust the zero-modes of the generators and
are dependent
in case $\mathrm{rank}\left\|  \sum_{k}\hat{\mathcal{Z}}_{k}{}_{\mathcal{A}%
_{1}}^{\mathcal{A}_{0}}(\theta_{0})\left(
\partial_{\theta_{0}}\right) ^{k}\right\|
_{\Sigma}=\overline{M}_{0}-\overline{M}<\overline{M}_{1}$. As a
result, the dependence relations for eigenvectors that define a
\emph{general}
$L_{g}$-\emph{stage} \emph{reducible LSM} are given by%
\begin{align}
&  \hspace{-1em}\int d\theta^{\prime}{\hat{\mathcal{Z}}}_{\mathcal{A}_{s-1}%
}^{\mathcal{A}_{s-2}}(\theta_{s-2};{\theta}^{\prime}){\hat{\mathcal{Z}}%
}_{\mathcal{A}_{s}}^{\mathcal{A}_{s-1}}(\theta^{\prime};{\theta}_{s})=\int
d\theta^{\prime}{\Theta}_{J}(\theta^{\prime})\mathcal{L}_{\mathcal{A}_{s}%
}^{\mathcal{A}_{s-2}J}\left(  (\mathcal{A},\partial_{\theta}\mathcal{A}%
)(\theta_{s-2}),\theta_{s-2},\theta^{\prime};\theta_{s}\right)  ,\nonumber\\
&
\hspace{-1em}\overline{M}_{s-1}>\sum\nolimits_{k=0}^{s-1}(-1)^{k}\overline
{M}_{s-k-2}=\mathrm{rank}\left\|  \sum\nolimits_{k\geq0}\hat{\mathcal{Z}}_k%
{}_{\mathcal{A}_{s-1}}^{\mathcal{A}_{s-2}}(\theta_{s-2})\left(
\partial
_{\theta_{s-2}}\right)  ^{k}\right\|  _{\Sigma},\nonumber\\
&  \hspace{-1em}\overline{M}_{L_{g}}=\sum\nolimits_{k=0}^{L_{g}}(-1)^{k}%
\overline{M}_{L_{g}-k-1}=\mathrm{rank}\left\|  \sum\nolimits_{k\geq0}
\hat{\mathcal{Z}_k%
}{}_{\mathcal{A}_{L_{g}}}^{\mathcal{A}_{L_{g}-1}}(\theta_{L_{g}-1})\left(
\partial_{\theta_{L_{g}-1}}\right)  ^{k}\right\|  _{\Sigma},\nonumber\\
&  \hspace{-1em}\vec{\varepsilon}({\hat{\mathcal{Z}}}_{\mathcal{A}_{s+1}%
}^{\mathcal{A}_{s}})=\vec{\varepsilon}_{\mathcal{A}_{s}}+\vec{\varepsilon
}_{\mathcal{A}_{s+1}}+(1,0,1),\ {\hat{\mathcal{Z}}}_{\mathcal{A}_{0}%
}^{\mathcal{A}_{-1}}(\theta_{-1};{\theta}_{0})\equiv{\hat{\mathcal{R}}%
}_{\mathcal{A}_{0}}^{I}(\theta_{-1};{\theta}_{0}),\nonumber\\
&  \hspace{-1em}\mathcal{L}_{\mathcal{A}_{1}}^{\mathcal{A}_{-1}J}(\theta
_{-1},\theta^{\prime};\theta_{1})\equiv\mathcal{K}_{\mathcal{A}_{1}}%
^{IJ}(\theta_{-1},\theta^{\prime};\theta_{1})=-(-1)^{(\varepsilon
_{I}+1)(\varepsilon_{J}+1)}\mathcal{K}_{\mathcal{A}_{1}}^{JI}(\theta^{\prime
},\theta_{-1};\theta_{1}).\label{10}%
\end{align}
for $s=1,...,L_{g}$, $\mathcal{A}_{s}=1,...$, $M_{s}=M_{s+}+M_{s-}$,
$\overline{M}\equiv\overline{M}_{-1}$. For $L_{g}=0$, an LSM is an
irreducible \emph{general gauge theory}.

For an LSM of the form $S_{\mathrm{L}}(\theta)=T\left(
\partial_{\theta }\mathcal{A}(\theta)\right)  -S\left(
\mathcal{A}(\theta),\theta\right)$, the functions
${\Theta}_{I}(\theta)$, ${\Theta}_{I}(\theta) \in
\mathcal{M}_{\mathrm{CL}}\times\{\theta\}$, take the form of the
usual extremals ${\Theta}_{I}(\theta)=-S,_{I}\left(
{\mathcal{A}}(\theta
),\theta\right)  (-1)^{\varepsilon_{I}}=0$ for the functional $S_{0}%
(A)=S\left(  \mathcal{A}(0),0\right)  $ corresponding to $\theta=0$.
Condition (\ref{8}) and identities (\ref{9}) have a form usual
for $\theta=0$,
\begin{equation}
\mathrm{rank}\left\|  S,_{IJ}\left(  {\mathcal{A}}(\theta),\theta\right)
\right\|  _{\Sigma}=\overline{N}-\overline{M},\ \ S,_{I}\left(  \mathcal{A}%
(\theta),\theta\right)  {\mathcal{R}}_{0}{}_{\mathcal{A}_{0}}^{I}\left(
\mathcal{A}(\theta),\theta\right)  =0,\label{11}%
\end{equation}
with linearly-dependent (for $\overline{M}_{0}>\overline{M}$)
generators of \emph{special gauge transformations},
$\delta\mathcal{A}^{I}(\theta)=\mathcal{R}_{0}{}_{\mathcal{A}_{0}}^{I}\left(
\mathcal{A}(\theta)\right.$, $\left.\theta\right)
{\xi}_{0}^{\mathcal{A}_{0}}(\theta)$, which leave invariant only
$S(\theta)$, in contrast to $T(\theta)$. The dependence of
$\mathcal{R}_{0}{}_{\mathcal{A}_{0}}^{I}(\theta)$, as well as of
their zero-eigenvalue eigenvectors $\mathcal{Z}_{\mathcal{A}_{1}}%
^{\mathcal{A}_{0}}(\mathcal{A}(\theta),\theta)$, and so on, can also be
expressed by special relations of reducibility for $s=1,...,L_{g}$, namely,
\begin{align}
&  \mathcal{Z}_{\mathcal{A}_{s-1}}^{\mathcal{A}_{s-2}}(\mathcal{A}%
(\theta),\theta)\mathcal{Z}_{\mathcal{A}_{s}}^{\mathcal{A}_{s-1}}%
(\mathcal{A}(\theta),\theta)=S,_{J}(\theta)\mathcal{L}_{\mathcal{A}_{s}%
}^{\mathcal{A}_{s-2}J}(\mathcal{A}(\theta),\theta),\;\vec{\varepsilon
}(\mathcal{Z}_{\mathcal{A}_{s}}^{\mathcal{A}_{s-1}})=\vec{\varepsilon
}_{\mathcal{A}_{s-1}}+\vec{\varepsilon}_{\mathcal{A}_{s}},\nonumber\\
&  \mathcal{Z}_{\mathcal{A}_{0}}^{\mathcal{A}_{-1}}(\theta)\equiv
\mathcal{R}_{0}{}_{\mathcal{A}_{0}}^{I}(\theta),\;\mathcal{L}_{\mathcal{A}%
_{1}}^{\mathcal{A}_{-1}J}(\theta)\equiv\mathcal{K}_{\mathcal{A}_{1}}%
^{IJ}(\theta)=-(-1)^{\varepsilon_{I}\varepsilon_{J}}\mathcal{K}_{\mathcal{A}%
_{1}}^{JI}(\theta).\label{12}%
\end{align}
In case $\overline{M}_{L_{g}}=\sum_{k=0}^{L_{g}}(-1)^{k}\overline{M}%
_{L_{g}-k-1}=\mathrm{rank}\left\|  \mathcal{Z}{}_{\mathcal{A}_{L_{g}}%
}^{\mathcal{A}_{L_{g}-1}}\right\|  _{\Sigma}$, we shall refer to
(\ref{11}) and (\ref{12}) as a \emph{special gauge theory} of
$L_{g}$-stage reducibility. The gauge algebra of such a theory is
$\theta$-locally embedded into the gauge
algebra of a general gauge theory with the functional $Z[\mathcal{A}%
]=\partial_{\theta}(T(\theta)-S(\theta))$, which leads to a relation between
the eigenvectors,%
\begin{equation}
\hat{\mathcal{Z}}_{\mathcal{A}_{s}}^{\mathcal{A}_{s-1}}(\mathcal{A}%
(\theta_{s-1})\,,\theta_{s-1};\theta_{s})=-\delta(\theta_{s-1}-\theta
_{s})\mathcal{Z}_{\mathcal{A}_{s}}^{\mathcal{A}_{s-1}}(\mathcal{A}%
(\theta_{s-1}),\theta_{s-1})\text{,}\label{13}%
\end{equation}
and to a possible parametric dependence of structure functions
on $\partial_{\theta}\mathcal{A}^{I}\left( \theta\right)$.
For special gauge theories in the Hamiltonian formulation,
definitions (\ref{11}) and (\ref{12}) retain their form; while
for general gauge theories of $L_{g}$-stage reducibility
definitions (\ref{8}) and (\ref{9}) are transformed:
\begin{equation}
{\hat{\mathcal{Z}}}_{\mathrm{H}}\,_{\mathcal{A}_{s}}^{\mathcal{A}_{s-1}%
}\left(  \Gamma_{k}(\theta_{s-1}),\theta_{s-1};\theta_{s}\right)
={\hat{\mathcal{Z}}}_{\mathcal{A}_{s}}^{\mathcal{A}_{s-1}}\left(
\mathcal{A}(\theta_{s-1}),\partial_{\theta_{s-1}}\mathcal{A}(\Gamma_{k}%
(\theta_{s-1}),\theta_{s-1}),\theta_{s-1};\theta_{s}\right)
\,,\;s=0,...,L_{g}\,.\label{14}%
\end{equation}

The extension of a usual field theory to a $\theta$-local LSM
permits one to apply Noether's first theorem \cite{Noether} to the
invariance of the density $d\theta S_{\mathrm{L}}(\theta)$ under
global $\theta$-translations, as symmetry transformations
$(\mathcal{A}^{I},z^{M},\theta)\rightarrow
(\mathcal{A}^{I},z^{M},\theta+\mu)$. One readily checks that
the function $S_{E}(\theta)$, and thus the action $S_{\mathrm{H}}(\theta)$,
identical with the former in terms of
$\Pi T^{\ast}\mathcal{M}_{\mathrm{CL}}$-coordinates,
\begin{equation}
S_{E}\left(  (\mathcal{A},\partial_{\theta}\mathcal{A})(\theta),\theta\right)
\equiv\frac{\partial S_{\mathrm{L}}(\theta)}{\partial(\partial_{\theta}%
^{r}\mathcal{A}^{I}(\theta))}\partial_{\theta}^{r}\mathcal{A}^{I}%
(\theta)-S_{\mathrm{L}}(\theta)\label{15},%
\end{equation}
are respective LS and HS integrals of motion, namely, quantities preserved
by the $\theta$-evolution, in case
\begin{eqnarray}
\left.
\frac{\partial}{\partial\theta}S_{\mathrm{L}}(\theta)+2(\partial
_{\theta}U)(\theta)S_{\mathrm{L}}(\theta)\right|  _{\mathcal{L}_{I}%
^{l}S_{\mathrm{L}}=0}=0, \ \
\frac{\partial}{\partial\theta}S_{\mathrm{H}}(\theta)-\left(
S_{\mathrm{H}}(\theta),S_{\mathrm{H}}(\theta)\right)  _{\theta}
=0. \label{16}
\end{eqnarray}
Provided that $S_{\mathrm{H}}(\theta)$ or $S_{\mathrm{L}}(\theta)$ do
not depend on $\theta$ explicitly, (\ref{16}) yields the
equations
$(S_{\mathrm{H}}(\theta)$, $S_{\mathrm{H}}(\theta))  {_{\theta}%
}=0$ or $\left.
(\partial_{\theta}U)(\theta)S_{\mathrm{L}}(\theta)\right|
_{\hat{\mathcal{A}}(\theta)}=0$, having no analogy in a $t$-local
field theory and implying the condition \cite{BV} that
$S_{\mathrm{H}}(\theta)$ or $S_{\mathrm{L}}(\theta)$ be proper,
however for an LSM on the classical level. Then the
$\theta$-superfield integrability\footnote{We use the notion of
$\theta$-superfield integrability by analogy with \cite{BatalinBeringDamgaard2}.}
of the HS in (\ref{3}) is implied by
the properties of the antibracket, in particular, the Jacobi identity,%
\begin{equation}
(\partial_{\theta}^{r})^{2}\Gamma_{k}^{P}(\theta)={\frac{1}{2}}\Bigl(
{\Gamma_{k}^{P}(\theta),}\Bigl(  {S_{\mathrm{H}}(\Gamma_{k}(\theta
)),S_{\mathrm{H}}(\Gamma_{k}(\theta))}\Bigr)  {_{\theta}}\Bigr)  {_{\theta}%
}=0.\label{19}%
\end{equation}
This yields a $\theta$-translation formula and the nilpotency of
the BRST-like generator $\check{s}^{l}(\theta)$ of $\theta$-shifts
along the $(\varepsilon_{P},\varepsilon)$-odd vector field
$\mathbf{Q}(\theta) = (S_{\mathrm{H}%
}(\theta),\,\cdot\,)_{\theta}$
\begin{equation}
\left.  \delta_{\mu}\mathcal{F}(\theta)\right|  _{\hat{\Gamma}_{k}(\theta
)}=\mu\left[  {\partial}/{\partial\theta}-\mathbf{Q}(\theta)\right]
\mathcal{F}(\theta)\equiv\mu\check{s}^{l}(\theta)\mathcal{F}(\theta
).\label{20}%
\end{equation}
Depending on additional properties (see Section 3) of a gauge theory,
we shall suppose
\begin{equation}
\Delta^{k}(\theta)S_{\mathrm{H}}(\theta)=0,\;\Delta^{k}(\theta)\equiv
\frac{1}{2}(-1)^{\varepsilon(\Gamma^{Q})}\omega_{QP}^{k}(\theta)\left(
\Gamma_{k}^{P}(\theta),\left(  \Gamma_{k}^{Q}(\theta),\,\cdot\,\right)
_{\theta}\right)  _{\theta},\label{21}%
\end{equation}
which is equivalent to a vanishing antisymplectic  divergence of
$\mathbf{Q}(\theta)$, $\left({\partial_{r}}/{\partial\Gamma_{k}^{P}%
(\theta)}\right)\mathbf{Q}(\theta) =
2\Delta^{k}(\theta)S_{\mathrm{H}}(\theta )=0$, which holds
trivially for its symplectic counterpart.
The \emph{Hamiltonian
master equation} $( S_{\mathrm{H}}(\theta
)$, $S_{\mathrm{H}}(\theta))_{\theta}=0$ for $({\partial}/%
{\partial\theta})S_{\mathrm{H}}(\theta)=0$ explains the
interpretation of the equivalent equation in (\ref{16}), for
$({\partial}/{\partial\theta })S_{\mathrm{L}}(\theta)=0$, $\left.
(\partial_{\theta}U)(\theta )S_{\mathrm{L}}(\theta)\right|
_{\mathcal{L}_{I}^{l}S_{\mathrm{L}}=0} = 0$, as a \emph{Lagrangian
master equation}.

\section{Superfield Quantization}
\subsection{Superfield Construction of a Local Quantum Action}

Here, the reducibility relations of a \emph{restricted }special
LSM are transformed into new gauge transformations for ghost
superfields. Along with the gauge transformations of
$\mathcal{A}^{i}(\theta)$ extracted from
$\mathcal{A}^{I}(\theta)$, the new gauge transformations imply a
Hamiltonian system related to an initial restricted HS
and leading to quantum and gauge-fixing actions subject
to respective $\theta$-local master equations.
With the standard distribution of ghost number \cite{BV}
$\mathrm{gh}(\mathcal{A}_{I}^{\ast}%
)=-1-\mathrm{gh}(\mathcal{A}^{I})=-1$ and the choice $\mathrm{gh}%
(\theta,\partial_{\theta})=(-1,1)$ implying the absence of ghosts among
$\mathcal{A}^{I}$, $(\varepsilon_{P})_{I}=0$, the quantization
is firstly given by
\begin{equation}
\left(  \mathrm{gh},{\partial}/{\partial\theta}\right)  S_{\mathrm{H}%
(\mathrm{L})}(\theta)=(0,0).\label{22}%
\end{equation}
Assuming the existence in $S_{\mathrm{H}(\mathrm{L})}(\theta)$ of
a potential term, $S(\mathcal{A}(\theta),0)=\mathcal{S}(
\mathcal{A}(\theta))$, a solution of (\ref{22}) extracts a
usual gauge theory with a classical action $S_{0}(A)$, where
$A^{i}$ are extended to $\mathcal{A}^{i}(\theta)$. The generalized HS
in (\ref{3}) then transforms into a $\theta$-integrable system on
$\Pi T^{\ast}\mathcal{M}_{\mathrm{cl}}=\{\Gamma^p_{\mathrm{cl}}(\theta)\}= \{(\mathcal{A}^{i},\mathcal{A}%
_{i}^{\ast})(\theta)\}$ with $\Theta_{i}(\mathcal{A}%
(\theta)) \in \mathcal{M}_{\mathrm{cl}}$,
\begin{equation}
\partial_{\theta}^{r}\Gamma_{\mathrm{cl}}^{p}(\theta)=\left(  \Gamma
_{\mathrm{cl}}^{p}(\theta),S_{0}(\mathcal{A}(\theta))\right)  _{\theta
},\;\Theta_{i}(\mathcal{A}(\theta))=-(-1)^{\varepsilon_{i}}%
S_{0},_{i}(\mathcal{A}(\theta)).\label{23}%
\end{equation}
The restricted special gauge transformations $\delta\mathcal{A}^{i}%
(\theta)=\mathcal{R}_{0\alpha_{0}}^{i}\left(  \mathcal{A}(\theta)\right)
{\xi}_{0}^{\alpha_{0}}(\theta)$, $\vec{\varepsilon}({\xi}_{0}^{\alpha_{0}%
}(\theta))=\vec{\varepsilon}_{\alpha_{0}}$, $(\varepsilon_{P})_{\alpha
_{0}}=0$, are embedded by $\xi_{0}^{\alpha_{0}}(\theta)=d\tilde{\xi}%
_{0}^{\alpha_{0}}(\theta)=\mathcal{C}^{\alpha_{0}}(\theta)d\theta$,
$\alpha_{0}=1,...$, $m_{0}=m_{0-}+m_{0+}$, into an HS of $2n$
equations for $\Gamma_{\mathrm{cl}}^{p}(\theta)$ with the
Hamiltonian
$S_{1}^{0}(\Gamma_{\mathrm{cl}},C_{0})(\theta)=(\mathcal{A}_{i}^{\ast
}\mathcal{R}_{0}{}_{\alpha_{0}}^{i}(\mathcal{A})\mathcal{C}^{\alpha_{0}%
})(\theta)$. A union of this system with the HS in (\ref{23}),
extended to $2(n+m_{0})$ equations, has the form%
\begin{equation}
\partial_{\theta}^{r}\Gamma_{\lbrack0]}^{p_{[0]}}(\theta)=\left(
\Gamma_{\lbrack0]}^{p_{[0]}}(\theta),S_{[1]}^{0}(\theta)\right)  _{\theta
},\;S_{[1]}^{0}(\theta)=(S_{0}+S_{1}^{0})(\theta),\;\Gamma_{\lbrack
0]}^{p_{[0]}}\equiv(\Gamma_{\mathrm{cl}}^{p},\Gamma_{0}^{p_{0}}),\;\Gamma
_{0}^{p_{0}}\equiv(\mathcal{C}^{\alpha_{0}},\mathcal{C}_{\alpha_{0}}^{\ast
}).\label{24}%
\end{equation}
Due to (\ref{12}), $S_{1}^{0}(\theta)$ is invariant, modulo $S_{0}%
,_{i}(\theta)$, under special gauge transformations for the ghost superfields
$\mathcal{C}^{\alpha_{0}}(\theta)$ with arbitrary functions $\xi_{1}%
^{\alpha_{1}}(\theta)$, $(\varepsilon_{P})_{\alpha_{1}}=0$,
defined in $\mathcal{M}$:%
\begin{equation}
\delta\mathcal{C}^{\alpha_{0}}(\theta)=\mathcal{Z}_{\alpha_{1}}^{\alpha_{0}%
}(\mathcal{A}(\theta))\xi_{1}^{\alpha_{1}}(\theta),\;(\vec{\varepsilon
},\mathrm{gh})\xi_{1}^{\alpha_{1}}(\theta)=\left(  \vec{\varepsilon}%
_{\alpha_{1}}+(1,0,1),1\right)  .\label{25}%
\end{equation}
In (\ref{25}), we now choose $\xi_{1}^{\alpha_{1}}(\theta)=d\tilde{\xi}%
_{1}^{\alpha_{1}}(\theta)=\mathcal{C}^{\alpha_{1}}(\theta)d\theta$,
$\alpha_{1}=1,...,m_{1}$, and extend the $m_{0}$
equations for $\mathcal{C}^{\alpha_{0}}(\theta)$ to an HS
of $2m_{0}$ equations with the Hamiltonian $S_{1}^{1}(\mathcal{A}%
,\mathcal{C}_{0}^{\ast},\mathcal{C}_{1})(\theta)=(\mathcal{C}_{\alpha_{0}%
}^{\ast}\mathcal{Z}_{\alpha_{1}}^{\alpha_{0}}(\mathcal{A})\mathcal{C}%
^{\alpha_{1}})(\theta)$. We then obtain a system (\ref{24}) for
$\partial_{\theta}^{r}\Gamma_{0}^{p_{0}}(\theta)$.
The extension of a union of the latter HS with eqs. (\ref{24}) is
formally identical to (\ref{24}) with the replacement%
\[
(\Gamma_{\lbrack0]}^{p_{[0]}},S_{[1]}^{0})\rightarrow(\Gamma_{\lbrack
1]}^{p_{[1]}},S_{[1]}^{1}):\;\left\{  \Gamma_{\lbrack1]}^{p_{[1]}}%
=(\Gamma_{\lbrack0]}^{p_{[0]}},\Gamma_{1}^{p_{1}}),\;\Gamma_{1}^{p_{1}%
}=(\mathcal{C}^{\alpha_{1}},\mathcal{C}_{\alpha_{1}}^{\ast}),\;S_{[1]}%
^{1}=S_{[1]}^{0}+S_{1}^{1}\right\}  .
\]
For an $L$-stage-reducible restricted LSM at the $s$-th step,
$0<s\leq L$, $\Gamma_{\mathrm{cl}}^{p}\equiv\Gamma_{-1}^{p_{-1}}$,
an iteration corresponding to reformulated special gauge
transformations for
$\mathcal{C}^{\alpha_{0}},...,\mathcal{C}^{\alpha_{s-2}}$ implied
by (possibly) enhanced\footnote{From
$\mathrm{gh}(\mathcal{A}^{I})=0$ in eqs.
(\ref{22}), with $(\varepsilon_{P})_{\mathcal{A}_{s}}=(\varepsilon_{P})_{I}%
=0$, $s=0,...,L_{g}$, it follows that $\overline{m}$, $\overline{m}_{s}$
may be both larger and smaller than $\overline{M}$, $\overline{M}_{s}$,
in contrast to $\overline{n}$, $\overline{N}$. In
fact, a restricted LSM may possess additional gauge symmetries.
Thus, we suppose that (possibly) enhanced sets of restricted
functions $\mathcal{R}_{0}{}_{\alpha_{0}}^{i}(\theta)$,
$\mathcal{Z}_{\alpha_{s}}^{\alpha_{s-1}}(\theta)$ exhaust, on the
surface $S_{0},_{i}(\theta)=0$, the zero-modes of both
$S_{0},_{ij}(\theta)$\thinspace
and\thinspace$\mathcal{Z}_{\alpha_{s-1}}^{\alpha_{s-2}}(\theta)$,
respectively. At\thinspace the\thinspace final stage of
reducibility for a restricted model, the above implies that $L\neq
L_{g}$.} formulae (\ref{12}) yields invariance transformations
for $S_{1}^{s-1}(\theta)$, modulo
$S_{0},_{i}(\theta)$,%
\begin{align}
&  \delta\mathcal{C}^{\alpha_{s-1}}(\theta)=\mathcal{Z}_{\alpha_{s}}%
^{\alpha_{s-1}}(\mathcal{A}(\theta))\xi_{s}^{\alpha_{s}}(\theta),\;(\vec
{\varepsilon},\mathrm{gh})\xi_{s}^{\alpha_{s}}(\theta)=(\vec{\varepsilon
}_{\alpha_{s}}+s(1,0,1),s),\;(\varepsilon_{P})_{\alpha_{s}}=0,\nonumber\\
&  S_{1}^{s-1}(\theta)=(\mathcal{C}_{\alpha_{s-2}}^{\ast}\mathcal{Z}%
_{\alpha_{s-1}}^{\alpha_{s-2}}(\mathcal{A})\mathcal{C}^{\alpha_{s-1}}%
)(\theta),\;\left(  \mathrm{gh},{\partial}/{\partial\theta}\right)
S_{1}^{s-1}(\theta)=(0,0).\label{26}%
\end{align}
The replacement $\xi_{s}^{\alpha_{s}}(\theta)=d\tilde{\xi}_{s}^{\alpha_{s}%
}(\theta)=\mathcal{C}^{\alpha_{s}}(\theta)d\theta$,
$\alpha_{s}=1,...$, $m_{s}=m_{s-}+m_{s+}$, transforms (\ref{26})
into $m_{s-1}$ equations for
$\mathcal{C}^{\alpha_{s-1}}(\theta)$, enlarged by an introduction
of $\mathcal{C}_{\alpha_{s-1}}^{\ast}(\theta)$ to the following HS:
\begin{equation}
\partial_{\theta}^{r}\Gamma_{s-1}^{p_{s-1}}(\theta)=\left(  \Gamma
_{s-1}^{p_{s-1}}(\theta),S_{1}^{s}(\theta)\right)  _{\theta},\;S_{1}%
^{s}(\theta)=(\mathcal{C}_{\alpha_{s-1}}^{\ast}\mathcal{Z}_{\alpha_{s}%
}^{\alpha_{s-1}}(\mathcal{A})\mathcal{C}^{\alpha_{s}})(\theta),\;\Gamma
_{s-1}^{p_{s-1}}=(\mathcal{C}^{\alpha_{s-1}},\mathcal{C}_{\alpha_{s-1}}^{\ast
}).\label{27}%
\end{equation}
Making a combination of (\ref{27}) with an HS of the same form,
however with $\partial_{\theta}^{r}\Gamma_{\lbrack
s-1]}^{p_{[s-1]}}(\theta)$ and with the Hamiltonian
$S_{[1]}^{s-1}(\theta)=(S_{0}+\sum_{r=0}^{s-1}S_{1}^{r})(\theta)$,
and presenting the result for $2\left(
n+\sum_{r=0}^{s}m_{r}\right) $ equations with
$S_{[1]}^{s}(\theta)=(S_{[1]}^{s-1}+S_{1}^{s})(\theta)$, we
obtain, by induction,%
\begin{equation}
\partial_{\theta}^{r}\Gamma_{\lbrack L]}^{p_{[L]}}(\theta)=\left(
\Gamma_{\lbrack L]}^{p_{[L]}}(\theta),S_{[1]}^{L}(\theta)\right)  _{\theta
}\,,\;S_{[1]}^{L}(\theta)=S_{0}(\mathcal{A}(\theta))+\sum\nolimits_{s=0}%
^{L}(\mathcal{C}_{\alpha_{s-1}}^{\ast}\mathcal{Z}_{\alpha_{s}}^{\alpha_{s-1}%
}(\mathcal{A})\mathcal{C}^{\alpha_{s}})(\theta).\label{28}%
\end{equation}
With the antibracket extended to $\Pi T^{\ast}\mathcal{M}_{k}$,
the function $S_{[1]}^{L}(\theta)$ is a proper solution
\cite{BV} of the classical master equation with accuracy up to
$O(C^{\alpha_{s}})$, modulo $S_{0},_{i}(\theta)$. An integrability
of the HS in (\ref{28}) is implied by a deformation of
$S_{[1]}^{L}(\theta)$ in powers of both
$\Phi_{A_{k}}^{\ast}(\theta)$ and
$\mathcal{C}^{\alpha_{s}}(\theta)$, by virtue of a superfield
counterpart of the existence theorem
\cite{BV} for the classical master equation in the minimal sector:%
\begin{equation}
\left(  S_{\mathrm{H};k}(\Gamma_{k}(\theta)),S_{\mathrm{H};k}(\Gamma
_{k}(\theta))\right)  _{\theta}=0,\;\left(  \vec{\varepsilon},\mathrm{gh}%
,{\partial}/{\partial\theta}\right)  S_{\mathrm{H};k}(\Gamma_{k}%
(\theta))=\left(  \vec{0},0,0\right)  ,\;k=\mathrm{min}.\label{29}%
\end{equation}
The construction of $S_{\mathrm{H};\mathrm{min}}(\theta)$ is
a superfield analogue of the Koszul--Tate complex resolution
\cite{KoszulTate}. Assuming now $k=\mathrm{ext}$, we remind that
the enlargement of $S_{\mathrm{H}%
;\mathrm{min}}(\theta)$ to $S_{\mathrm{H};k}(\Gamma_{k}(\theta))$,
$S_{\mathrm{H};k}(\theta) = S_{\mathrm{H};\mathrm{min}}(\theta)+\sum_{s=0}%
^{L}\sum_{s^{\prime}=0}^{s}(\mathcal{C}_{s^{\prime}{}\alpha_{s}}^{\ast
}\mathcal{B}_{s^{\prime}}^{\alpha_{s}})(\theta)$, being a proper solution
in $\Pi T^{\ast}\mathcal{M}_{k}$,
with a deformation in the Planck constant $\hbar$, determines a quantum action
$S_{\mathrm{H}}^{\Psi }(\Gamma(\theta),\hbar)$, e.g., in case of
an Abelian hypergauge,%
\begin{eqnarray}
\left[\Gamma_{k}^{p_{k}}(\theta)\mapsto{\Gamma^{\prime}}_{k}^{p_{k}}%
(\theta)=\left(  \Phi^{A_{k}}(\theta),{\Phi}_{A_{k}}^{\ast}(\theta
)-\frac{\partial\Psi(\Phi(\theta))}{\partial\Phi^{A_{k}}(\theta)}\right)\right]
\Rightarrow
S_{\mathrm{H}}^{\Psi}(\Gamma(\theta),\hbar)=e^{\left({\Psi}(\Phi(\theta)),\ \cdot\  \right) _{\theta}%
}S_{\mathrm{H};k}(\Gamma_{k}(\theta),\hbar).\label{30}%
\end{eqnarray}
The functions
$(S_{\mathrm{H}}^{\Psi},S_{\mathrm{H};k})(\theta,\hbar)$ obey
equations (\ref{21}), (\ref{29}) provided that the
$\hbar$-deformation of $S_{\mathrm{H};\mathrm{min}}(\theta)$ is
their solution. Such equations are known to ensure an
integrability of the non-equivalent HS constructed from
$S_{\mathrm{H}}^{\Psi}$, $S_{\mathrm{H};k}$, as well as the
anticanonical [respecting the volume
$dV_{k}(\theta)=\prod_{p_{k}}d\Gamma _{k}^{p_{k}}(\theta)$] nature
of transformations (\ref{30}), related to a $\theta $-shift
by a constant $\mu$ along the corresponding HS solutions.
At the same time, the quantum master equation%
\begin{equation}
\Delta^{k}(\theta)\exp\left[  ({i}/{\hbar})E(\theta,\hbar)\right]
=0,\;E\in\{S_{\mathrm{H}}^{\Psi},S_{\mathrm{H};k}\}\label{31}%
\end{equation}
introduces a non-integrable HS, with the corresponding
anticanonical transformation preserving
$d\hat{V}_{k}(\theta)=\exp\left[  \left( i/\hbar\right)
E(\theta,\hbar)\right]  dV_{k}(\theta)$. The vector field related
to the latter HS determines, for
$\partial_{\theta}^{r}{\Phi}_{A_{k}}^{\ast}(\theta)=0$,
a $\theta$-local generator $\tilde{s}^{l(\Psi)}%
(\theta)$ of BRST transformations crucial for the BV formalism (in
case $\theta = 0$).

\subsection{BV--BFV Duality}

We now propose a dual description for an LSM. Namely, an embedding of a
restricted LSM gauge algebra with $S_{\mathrm{H};\mathrm{min}}(\theta)$
and equations (\ref{29}) into the gauge algebra of a
general gauge theory in Lagrangian formalism
(\ref{8})--(\ref{12}) can be realized by dual counterparts, with
the opposite ($\varepsilon_{P},\varepsilon$)-parity, of the action
and antibracket, following, in part, Refs.
\cite{AlexandrovKontsevichSchwarzZaboronsky,GrigorievDamgaard}.
For this purpose, consider the functional%
\[
Z_{k}[\Gamma_{k}]=-\partial_{\theta}S_{\mathrm{H};k}(\theta)\,,\;(\vec
{\varepsilon},\mathrm{gh})Z_{k}=((1,0,1),1)
\]
in $\Pi T(\Pi
T^{\ast}\mathcal{M}_{k})=\{(\Gamma_{k}^{p_{k}},\partial_{\theta
}\Gamma_{k}^{p_{k}})(\theta),k=\mathrm{min}\}$, with a symplectic
and odd Poisson structures which define an even
functional $\{\,\cdot\,,\,\cdot\,\}$ with canonical pairs $\{(\Phi_{k}^{A_{k}%
},\partial_{\theta}\Phi_{A_{k}}^{\ast})$, $(\partial_{\theta}\Phi_{k}^{A_{k}%
}, \Phi_{A_{k}}^{\ast})$ $\}(\theta)$ and a $\theta$-local odd
Poisson bracket
$(\,\cdot\,,\,\cdot\,)_{\theta}^{(\Gamma_{k},\partial_{\theta
}\Gamma_{k})}$. The latter bracket acts on
$C^{\infty}(\Pi T(\Pi T^{\ast}\mathcal{M}_{k})$ $\times\theta)$
and provides a lifting of $(\,\cdot\,,\,\cdot\,)_{\theta}$ defined
in $\Pi T^{\ast}\mathcal{M}_{k}$. For
any $F_{\mathrm{t}}[\Gamma_{k}]=\partial_{\theta}\mathcal{F}%
_{\mathrm{t}}\left(  (\Gamma_{k},\partial_{\theta}\Gamma_{k})(\theta
),\theta\right)  $, $\mathrm{t}=1,2$, there holds a correspondence between
the Poisson brackets of the opposite Grassmann gradings:%
\begin{align}
&  \hspace{-1em}\left\{  F_{1},F_{2}\right\}  =\int d\theta\left[
\frac{\delta F_{1}}{\delta\Phi^{A_{k}}(\theta)}\frac{\delta F_{2}}{\delta
\Phi_{A_{k}}^{\ast}(\theta)}-\frac{\delta_{r}F_{1}}{\delta\Phi_{A_{k}}^{\ast
}(\theta)}\frac{\delta_{l}F_{2}}{\delta\Phi^{A_{k}}(\theta)}\right]  =\int
d\theta(\mathcal{F}_{1}(\theta),\mathcal{F}_{2}(\theta))_{\theta}^{(\Gamma
_{k},\partial_{\theta}\Gamma_{k})},\nonumber\\
&  \hspace{-1em}(\mathcal{F}_{1}(\theta),\mathcal{F}_{2}(\theta))_{\theta
}^{(\Gamma_{k},\partial_{\theta}\Gamma_{k})}\equiv\left[  \left(
\mathcal{L}_{A_{k}}\mathcal{F}_{1}\right)  \mathcal{L}^{\ast A_{k}}%
\mathcal{F}_{2}-(\mathcal{L}_{r}^{\ast A_{k}}\mathcal{F}_{1})\mathcal{L}%
_{A_{k}}^{l}\mathcal{F}_{2}\right]  (\theta).\label{33}%
\end{align}
Here, for instance, the Euler--Lagrange superfield derivative with respect to
$\Phi_{A_{k}}^{\ast}(\theta)$, for a fixed $\theta$, is given by $\mathcal{L}%
^{\ast A_{k}}(\theta)=\partial/\partial\Phi_{A_{k}}^{\ast}(\theta
)-(-1)^{\varepsilon_{A_{k}}+1}\partial_{\theta}\cdot\partial/\partial\left(
\partial_{\theta}\Phi_{A_{k}}^{\ast}(\theta)\right)  $.

The functional $Z_{k}$, for $k = \mathrm{min}$, is nilpotent by
construction: $\left\{  Z_{k},Z_{k}\right\} =
\partial_{\theta}\left(S_{\mathrm{H};k}(\theta
),S_{\mathrm{H};k}(\theta)\right) _{\theta}=0$. The absence of
a time coordinate implies that $Z_{k}$ formally corresponds to a
BRST charge for a dynamical system with first-class constraints
\cite{BFV}. In fact, after identifying $(\Gamma_{k},\partial
_{\theta}\Gamma_{k})(0)$ with phase-space coordinates (of the
minimal sector) canonical with respect to the
$(\varepsilon_{P},\varepsilon)$-even BFV bracket
for a first-class constrained system of $(L+1)$-stage reducibility,%
\begin{align}
&  (q^{i},p_{i})=(\mathcal{A}^{i},\partial_{\theta}\mathcal{A}_{i}^{\ast
})(0),\;\left(  C^{A_{s}},\mathcal{P}_{A_{s}}\right)  =\left(  (\partial
_{\theta}^{r}\mathcal{C}^{\alpha_{s-1}},\mathcal{C}^{\alpha_{s}}%
),(\mathcal{C}_{\alpha_{s-1}}^{\ast},\partial_{\theta}\mathcal{C}_{\alpha_{s}%
}^{\ast})\right)  (0),\nonumber\\
&  A_{s}=(\alpha_{s-1},\alpha_{s}),\ s=0,...,L,\;\left(  C^{A_{L+1}%
},\mathcal{P}_{A_{L+1}}\right)  =\left(  (\partial_{\theta}^{r}\mathcal{C}%
^{\alpha_{L}}, 0),(\mathcal{C}_{\alpha_{L}}^{\ast},0)\right)  (0),\label{34}%
\end{align}
the functional $Z_{k}$ acquires the form%
\begin{equation}
Z_{k}[\Gamma_{k}]=T_{A_{0}}(q,p)C^{A_{0}}+\sum\limits_{s=1}^{L+1}%
\mathcal{P}_{A_{s-1}}Z_{A_{s}}^{A_{s-1}}(q)C^{A_{s}}+O(C^{2}).\label{35}%
\end{equation}
The structure functions of the initial $L$-stage reducible restricted LSM in
the enhanced eqs. (\ref{9}) determine $T_{A_{0}}(q,p)$ and
a set of $(L+1)$-stage reducible eigenvectors $Z_{A_{s}}^{A_{s-1}}(q)$:%
\begin{align}
&  \hspace{-2em}T_{A_{0}}(q,p)=\left(  S_{0},_{i}(q),-p_{i}\mathcal{R}_{0}%
{}_{\alpha_{0}}^{i}(q)\right) ,\;Z_{A_{s}}^{A_{s-1}}(q) =
\mathrm{diag}\left( \mathcal{Z}_{\alpha_{s-1}}^{\alpha_{s-2}},
\left[1
- \delta^{\alpha_{s}}_{\alpha_{L+1}}\right]\mathcal{Z}_{\alpha_{s}}%
^{\alpha_{s-1}}\right)(q), \label{36}\\
&  \hspace{-2em}Z_{A_{s-1}}^{A_{s-2}}Z_{A_{s}}^{A_{s-1}}=T_{B_{0}}L_{A_{s}%
}^{A_{s-2}B_{0}}(q,p),\;s=1,...,\;L+1,\;Z_{A_{0}}^{A_{-1}}\equiv T_{A_{0}%
},\ L_{A_{s}}^{A_{s-2}\beta_{0}}=0,\nonumber\\
&  \hspace{-2em}L_{A_{s}}^{A_{s-2}j}=\mathrm{diag\,}\left(  \mathcal{L}%
_{\alpha_{s-1}}^{\alpha_{s-3}j},\;\mathcal{L}_{\alpha_{s}}^{\alpha_{s-2}%
j}\right)  ,\;\mathcal{L}_{\alpha_{0}}^{\alpha_{-2}j}=\mathcal{L}%
_{\alpha_{L+1}}^{\alpha_{L-1}j}=0,\;\mathcal{L}_{\alpha_{1}}^{\alpha_{-1}%
j}(q,p)=(-1)^{\varepsilon_{j}+1}p_{i}\mathcal{K}_{\alpha_{1}}^{ji}%
(q).\label{37}%
\end{align}
Formulae (\ref{33})--(\ref{37}) generalize to any reducible
theories the dual description (proposed for
$\varepsilon_{i}=\varepsilon_{\alpha_{0}}=L=0$) of a quantum
action in the minimal sector via a nilpotent BRST charge
in the minimal sector \cite{GrigorievDamgaard}. One can show that
the corresponding dual description in terms of the extended
variables of the BV and BFV methods yields the only
possible embedding of $Z_{k}[\Gamma_k]$, $k = \mathrm{ext}$, and
$\Pi T(\Pi T^{\ast}\mathcal{M}_{k})$ into the BRST charge and
total phase space of the BFV method.

A characteristic property of the duality problem is an equivalent
definition of the systems with the Hamiltonians $S_{\mathrm{H}%
}^{\Psi}(\Gamma(\theta),\hbar)$, $S_{\mathrm{H};k}(\theta)$, $k=\mathrm{min}$,
$\mathrm{ext}$, by means of dual fermionic functionals, $Z_{k}[\Gamma_{k}]$,
$Z^{\Psi}[\Gamma]=-\partial_{\theta}S_{\mathrm{H}}^{\Psi}(\Gamma(\theta
),\hbar)$, in terms of even Poisson brackets:%
\begin{equation}
\partial_{\theta}^{r}\Gamma^{p}(\theta)=\left(  \Gamma^{p}(\theta
),S_{\mathrm{H}}^{\Psi}(\Gamma(\theta),\hbar)\right)  _{\theta}=-\left\{
\Gamma^{p}(\theta),Z^{\Psi}[\Gamma]\right\}  .\label{38}%
\end{equation}
Thus, BRST transformations in a Lagrangian formalism with Abelian
hypergauges can be expressed in terms of a formal BRST charge
$Z^{\Psi}[\Gamma]$ related to $Z_{k}[\Gamma_{k}]$,
$k=\mathrm{ext}$, by a  canonical transformation with an even
phase, $F^{\Psi}[\Phi]$ = $\partial_{\theta}\Psi(\Phi(\theta))$,
$Z^{\Psi}[\Gamma]=\exp{\left\{F^{\Psi}%
,\ \cdot\  \right\}}Z_{k}[\Gamma_{k}]$.

\subsection{Local Quantization}
Let us define a generating functional of Green's functions $\mathsf{Z}%
(\theta)$ and an effective action $\mathsf{\Gamma}(\theta)$, using
an invariant description of super(anti)fields on a general
antisymplectic manifold. For this purpose, we use a choice of
Darboux coordinates $(\varphi,\varphi^{\ast})(\theta)$ consistent
with the properties of a quantum action. We suppose that a model
is described by a quantum action $W(\theta)\equiv W(\theta,\hbar)$
defined on an arbitrary (without connection) antisymplectic manifold
$\mathcal{N}$=$\{\Gamma^{p}(\theta)\}$, $\dim\mathcal{N}=\dim\Pi
T^{\ast}\mathcal{M}_{\mathrm{ext}}$, with a density function
$\rho(\Gamma(\theta))$  determining  an invariant volume element
$d\mu(\Gamma(\theta))$. A local antibracket and a nilpotent
second-order operator $\Delta^{\mathcal{N}}(\theta)$ are defined
with the help
of an odd Poisson bivector, $\omega^{pq}%
(\Gamma(\theta))=\left(  \Gamma^{p}(\theta),\ \Gamma^{q}(\theta)\right)
_{\theta}^{\mathcal{N}}$,%
\begin{equation}
\Delta
^{\mathcal{N}}(\theta)=\frac{1}{2}(-1)^{\varepsilon(\Gamma^{q})}\rho
^{-1}\omega_{qp}(\theta)\left(  \Gamma^{p}(\theta),\rho\left(
\Gamma ^{q}(\theta),\ \cdot\  \right)  _{\theta}^{\mathcal{N}}\right)
_{\theta
}^{\mathcal{N}}. \label{40}%
\end{equation}
In perturbation theory, a generating functional of Green's functions
$\mathsf{Z}\left(  (\partial_{\theta}\varphi^{\ast},\varphi^{\ast}%
,\partial_{\theta}\varphi,\mathcal{I})(\theta)\right)  \equiv\mathsf{Z}%
(\theta)$ can de defined as a path integral (for a fixed $\theta$) by
introducing on $\mathcal{N}$ some Darboux coordinates $\Gamma^{p}%
(\theta)=(\varphi^{a},\varphi_{a}^{\ast})(\theta)$ in a vicinity
of the stationary points of $ W(\theta)$ such that $\rho=1$ and
$\omega^{pq}(\theta)=\mathrm{antidiag}(-\delta_{b}^{a},\delta_{b}^{a})$.
The function
\begin{align}
\mathsf{Z}(\theta) &  =\int d\mu\left(  \tilde{\Gamma}(\theta)\right)
\,d\Lambda(\theta)\exp\left\{  \left(  i/\hbar\right)  \left[  {W}\left(
\tilde{\Gamma}(\theta),\hbar\right)  +\left.  X\left(  \left(  \tilde{\varphi
},\tilde{\varphi}^{\ast}-{\varphi}^{\ast},\Lambda,\Lambda^{\ast}\right)
(\theta),\hbar\right)  \right|  _{\Lambda^{\ast}=0}\right.  \right.
\nonumber\\
&  -\left.  \left.  ((\partial_{\theta}\varphi_{a}^{\ast})\tilde{\varphi}%
{}^{a}+\tilde{\varphi}{}_{a}^{\ast}\partial_{\theta}^{r}{\varphi}{}%
^{a}-\mathcal{I}_{a}\Lambda^{a})(\theta)\right]  \right\},\
d\mu\left({\Gamma}(\theta)\right) = \rho(\Gamma(\theta)) d \Gamma(\theta),  \label{41}%
\end{align}
depends on extended sources $(\partial_{\theta}\varphi_{a}^{\ast}%
,\partial_{\theta}^{r}{\varphi}{}^{a},\mathcal{I}_{a})(\theta)=(-J_{a}%
,\lambda^{a},I_{0a}+I_{1a}\theta)$ for
$(\varphi^{a},\varphi_{a}^{\ast},\Lambda^{a})(\theta)$ with the
properties
$(\vec{\varepsilon},\mathrm{gh})\partial_{\theta}\varphi_{a}^{\ast}%
=(\vec{\varepsilon},\mathrm{gh})\mathcal{I}_{a}+((1,0,1),1)=(\vec{\varepsilon
},-\mathrm{gh})\varphi^{a}$,   where $\Lambda
^{a}(\theta)=(\lambda_{0}^{a}+\lambda_{1}^{a}\theta)$ are
Lagrangian multipliers to independent non-Abelian hypergauges,
$G_{a}(\Gamma (\theta)),a=1,...,\;k=\dim_{+}\mathcal{N}$; see
\cite{BatalinTyutin}. The
functions $G_{a}(\Gamma(\theta))$, $(\vec{\varepsilon},\mathrm{gh})G_{a}%
=(\vec{\varepsilon},\mathrm{gh})\mathcal{I}_{a}$, determine a
boundary condition (when $\Lambda_{a}^{\ast}=\hbar=0$) for the
gauge-fixing action $X(\theta)=X\left(
(\Gamma,\Lambda,\Lambda^{\ast})(\theta),\hbar\right)  $ defined on
the direct sum $\mathcal{N}_{\mathrm{tot}}=\mathcal{N}\oplus\Pi
T^{\ast}\mathcal{K}$, where $\Pi
T^{\ast}\mathcal{K}$ = $\{(\Lambda^{a},\Lambda_{a}^{\ast
})(\theta)\}$. Hypergauges in involution
$(G_{a}(\theta),G_{b}(\theta
))_{\theta}^{\mathcal{N}}=G_{c}(\theta)U_{ab}^{c}(\Gamma(\theta))$
obey various unimodularity relations \cite{BatalinTyutin}
depending on equations with a solution $X(\theta)$, in terms of
the antibracket $(\,\cdot
\,,\,\cdot\,)_{\theta}=(\,\cdot\,,\,\cdot\,)_{\theta}^{\mathcal{N}}%
+(\,\cdot\,,\,\cdot\,)_{\theta}^{\mathcal{K}}$ and the operator $\Delta
(\theta)=(\Delta^{\mathcal{N}}+\Delta^{\mathcal{K}})(\theta)$,
trivially lifted from $\mathcal{N}$ to $\mathcal{N}_{\mathrm{tot}}$,%
\begin{equation}
1)\;(E(\theta),E(\theta))_{\theta}=0,\;\Delta(\theta)E(\theta
)=0;\ \ \ 2)\,\Delta(\theta)\exp\left[
({i}/{\hbar})E(\theta)\right]
=0,\;E\in\{W,X\}.\label{42}%
\end{equation}
The functions $G_{a}(\theta)$ solvable with respect to
$\varphi_{a}^{\ast }(\theta)$ determine a Lagrangian surface,
$\mathcal{Q}_{g}=\{({\varphi }^{\ast},
\Lambda)(\theta)\}\subset\mathcal{N}_{\mathrm{tot}}$, on which
$\left.  X(\theta)\right| _{\mathcal{Q}_{g}}$ is non-degenerate.
Then integration over $(\tilde{\varphi}^{\ast },\Lambda)(\theta)$
in eq. (\ref{41}) yields (when $\partial_{\theta}{\varphi
}{}^{a}=\mathcal{I}_{a}=0$) a function whose restriction to the
Lagrangian surface $\mathcal{Q}=\{{\varphi}(\theta)\}$ is also
non-degenerate.

Using the properties of $(W,X)(\theta)$, one can introduce an effective action
$\mathsf{\Gamma}(\theta)\equiv\mathsf{\Gamma}(\varphi,{\varphi}^{\ast
},\partial_{\theta}^{r}{\varphi},\mathcal{I})(\theta)$ by a Legendre
transformation of $\ln\mathsf{Z}(\theta)$ with respect to $\partial_{\theta
}{\varphi}_{a}^{\ast}(\theta)$,%
\begin{equation}
\mathsf{\Gamma}(\theta)=\left({\hbar}/{i}\right)\ln\mathsf{Z}(\theta)+\left(
(\partial_{\theta}\varphi_{a}^{\ast})\varphi^{a}\right)  (\theta
),\;\varphi^{a}(\theta)= {i}{\hbar}\frac{\partial_{l}\ln\mathsf{Z}%
(\theta)}{\partial(\partial_{\theta}\varphi_{a}^{\ast}(\theta))}\,.\label{43}%
\end{equation}
The properties of $(\mathsf{Z},\mathsf{\Gamma})(\theta)$ are
implied by a $\theta$-nonintegrable Hamiltonian-like system with
an arbitrary
$(\varepsilon_{P},\varepsilon)$-even $C^{\infty}(\mathcal{N}_{\mathrm{tot}}%
)$-function $R(\theta)=R\left(  (\tilde{\Gamma},\Lambda,\Lambda^{\ast}%
)(\theta),\hbar\right)  $, $\mathrm{gh}R(\theta)=0$, for
$T(\theta)=\exp\left[  \left(  i/\hbar\right)(W-X)(\theta)\right]$,
\begin{equation}
\partial_{\theta}^{r}\left(  \tilde{\Gamma}^{p},\Lambda^{a},\varphi_{a}^{\ast
},\Lambda_{a}^{\ast}\right)  (\theta)=-i\hbar T^{-1}(\theta)\left.
\left(  {\ \cdot\ ,T(\theta)R(\theta)}\right){_{\theta}}  \left(
\tilde{\Gamma}^{p},2\Lambda^{a}%
,0,0\right)  (\theta)\right|
_{\Lambda^{\ast}=0}.\label{44}%
\end{equation}
The integrand in (\ref{41}) is invariant (for
$\partial_{\theta}\varphi^{\ast
}=\partial_{\theta}{\varphi}=\mathcal{I}=0$)\thinspace under the
superfield \emph{\thinspace BRST \thinspace transformations}%
\begin{equation}
\tilde{\Gamma}_{\mathrm{tot}}(\theta)=(\tilde{\Gamma},\Lambda,\Lambda^{\ast
})(\theta)\rightarrow\left(  \tilde{\Gamma}_{\mathrm{tot}}+\delta_{\mu}%
\tilde{\Gamma}_{\mathrm{tot}}\right)  (\theta),\;\delta_{\mu}\tilde{\Gamma
}_{\mathrm{tot}}(\theta)=\left.  \left(  \partial_{\theta}^{r}\tilde{\Gamma
}_{\mathrm{tot}}\right)  \right|  _{\check{\Gamma}_{\mathrm{tot}}}%
\mu,\label{45}%
\end{equation}
related to a $\theta$-shift by a constant $\mu$ along an arbitrary
solution $\check{\Gamma}_{\mathrm{tot}}(\theta)$ of (\ref{44}),
for $R(\theta)=1$, with the arguments of $(W,X)(\theta)$ being
those in (\ref{41}).

The function
$\mathsf{Z}_{X}(\theta)\equiv\mathsf{Z}(0,\varphi^{\ast
},0,0)(\theta)$ is gauge-independent: it does not change if
$X(\theta)$ is replaced by $(X+\Delta X)(\theta)$ that obeys
equations (\ref{42}) for $X(\theta)$ and respects
nondegeneracy on $\mathcal{Q}_{g}$. This means that $\Delta
X(\theta)$\thinspace obeys\thinspace a\thinspace set\thinspace
of\thinspace linearized\thinspace equations\thinspace
with\thinspace a\thinspace nilpotent\thinspace operator
$Q_{j}(X)$,
\begin{equation}
Q_{j}(X)\Delta X(\theta)=0,\,\delta_{j1}\Delta(\theta)\Delta X(\theta
)=0;\;Q_{j}(X)=\mathrm{ad\,}X(\theta)-\delta_{j2}(i\hbar\Delta(\theta
)),\ j=1,2,\label{46}%
\end{equation}
where $j$ labels those equations in (\ref{42}) which are met by $X(\theta)$.
By analogy with the theorems
\cite{BatalinLavrovTyutin}, the fact that solutions $X(\theta)$ of
each system in (\ref{42}) are proper implies
that the cohomologies of $Q_{j}(X)$ on functions $f({\Gamma}_{\mathrm{tot}%
}(\theta))\in C^{\infty}(\mathcal{N}_{\mathrm{tot}})$ vanishing
for ${\Gamma }_{\mathrm{tot}}(\theta)=0$ are trivial. Thus, the
general solution of (\ref{46}) is given by a certain function
$\Delta Y\left( \theta\right)$, $\left.\Delta Y(\theta)\right|
_{{\Gamma}_{\mathrm{tot}}=0}=0$,
\begin{equation}
\Delta X(\theta)=Q_{j}(X)\Delta Y(\theta),\,\delta_{j1}\Delta(\theta)\Delta
Y(\theta)=0,\;\left(  \vec{\varepsilon},\mathrm{gh},{\partial}/{\partial
\theta}\right)  \Delta Y(\theta)=\left(  (1,0,1),-1,0\right).  \label{47}%
\end{equation}
 Making in $\mathsf{Z}_{X+\Delta X}(\theta)$ a change of
variables induced by a $\theta$-shift by a constant $\mu$, related
to (\ref{44}), and choosing $2R(\theta)\mu=\Delta Y(\theta)$, we
have $\mathsf{Z}_{X+\Delta X}(\theta)=\mathsf{Z}_{X}(\theta)$,
which implies the gauge-independence of the S-matrix, in view
of the equivalence theorem \cite{Tyutin2}.

Following Subsection 3.2, the stated properties of $\mathsf{Z}_{X}(\theta)$
can be independently derived from a Ha\-mil\-to\-ni\-an-like system in terms
of an even superfield Poisson bracket in general coordinates (see footnote
\thefootnote),
\begin{equation}
\partial_{\theta}^{r}\left( \tilde{\Gamma}^{p},\Lambda^{a},\varphi_{a}^{\ast
},\Lambda_{a}^{\ast}\right)  (\theta)=\left.  \left\{  \ \cdot\
,(Z^{X}+i\hbar
Z^{R})[\tilde{\Gamma}_{\mathrm{tot}}]-Z^{W}[\tilde{\Gamma}]\right\}
\left( \tilde{\Gamma}^{p},2\Lambda^{a},0,0\right)(\theta)\right|
_{\Lambda^{\ast}=0}
,\label{48}%
\end{equation}
where $Z^{E}[{\Gamma}_{\mathrm{tot}}]=-\partial_{\theta}E({\Gamma
}_{\mathrm{tot}}(\theta),\hbar),\;E\in\{W,X,R\}$. If
$(W,X)(\theta)$, obey the first system in (\ref{47}), then $Z^{W}$,
$Z^{X}$, playing the role of a usual and \emph{gauge-fixing}
BRST charge, are nilpotent with respect to the Poisson
bracket $\{\,\cdot\,,\,\cdot\,\}=\{\,\cdot\,,\,\cdot\,\}^{\Pi
T\mathcal{N}}+\{\,\cdot\,,\,\cdot\,\}^{\Pi T\mathcal{K}}$. Here,
the first
bracket is defined on any functionals over ${\Pi T\mathcal{N}}%
\times\{\theta\}$ in terms of a $\theta$-local extension
$(\,\cdot \,,\,\cdot\,)_{\theta}^{\Pi T\mathcal{N}}
= ((\mathcal{L}_{p}^{r} \,\cdot \,)
\omega ^{pq}(\Gamma(\theta))\mathcal{L}_{q}^{l}\,\cdot \,)$
of antibracket (\ref{33})
\begin{align}
&  \hspace{-1em}\left\{  F_{1},F_{2}\right\}  ^{\Pi T\mathcal{N}}\equiv\int
d\theta\frac{\delta_{r}F_{1}}{\delta\Gamma^{p}(\theta)}\omega^{pq}%
(\Gamma(\theta))\frac{\delta_{l}F_{2}}{\delta\Gamma^{q}(\theta)}%
=\partial_{\theta}(\mathcal{F}_{1}(\theta),\mathcal{F}_{2}(\theta))_{\theta
}^{\Pi T\mathcal{N}},\;F_{\mathrm{t}}[\Gamma]=\partial_{\theta}\mathcal{F}_{\mathrm{t}}%
((\Gamma,\partial_{\theta}\Gamma)(\theta),\theta),\label{49}%
\end{align}
where $\mathcal{L}_{p}^{l}(\theta)$ is the left-hand Euler--Lagrange
superfield derivative with respect to $\Gamma^{p}(\theta)$.

The functions $(\mathsf{Z},\mathsf{\Gamma})(\theta)$ obey the Ward
identities%
\begin{align}
&  \left\{  \partial_{\theta}\varphi_{a}^{\ast}(\theta)\frac{\partial_{l}%
}{\partial\varphi_{a}^{\ast}(\theta)}+\frac{i}{\hbar}\partial_{\theta}%
^{r}\varphi^{a}(\theta)\left[  \partial_{\theta}\varphi_{a}^{\ast}%
(\theta)-\frac{\partial
X}{\partial\tilde{\varphi}{}^{a}(\theta)}\left(
{i\hbar}\frac{\partial_{l}}{\partial(\partial_{\theta}\varphi^{\ast})}%
,{i\hbar}\frac{\partial_{r}}{\partial(\partial_{\theta}^{r}\varphi)}%
-\varphi^{\ast},\frac{\hbar}{i}\frac{\partial_{l}}{\partial\mathcal{I}%
},0\right)  \right]  \right.  \nonumber\\
&  \left.  +\frac{i}{\hbar}\mathcal{I}_{a}(\theta)\frac{\partial_{l}}%
{\partial\Lambda_{a}^{\ast}(\theta)}\left.  X\left(
{i\hbar}\frac{\partial
_{l}}{\partial(\partial_{\theta}\varphi^{\ast})},{i\hbar}\frac{\partial_{r}%
}{\partial(\partial_{\theta}^{r}\varphi)}-\varphi^{\ast},\frac{\hbar}%
{i}\frac{\partial_{l}}{\partial\mathcal{I}},\Lambda^{\ast}\right)
\right|
_{\Lambda_{a}^{\ast}=0}\right\}  \mathsf{Z}(\theta)=0,\label{50}\\
&
\mathcal{I}_{a}(\theta)\frac{\partial_{l}}{\partial\Lambda_{a}^{\ast
}(\theta)}\left.  X\left(
{\varphi}^{b}+i\hbar(\mathsf{\Gamma}^{\prime\prime
}{}^{-1})^{bc}\frac{\partial_{l}}{\partial\varphi^{c}},{i\hbar}\frac{\partial
_{r}}{\partial(\partial_{\theta}^{r}\varphi)}-\frac{\partial_{r}%
\mathsf{\Gamma}}{\partial(\partial_{\theta}^{r}\varphi)}-\varphi^{\ast
},\frac{\partial_{l}\mathsf{\Gamma}}{\partial\mathcal{I}}+\frac{\hbar}%
{i}\frac{\partial_{l}}{\partial\mathcal{I}},\Lambda^{\ast}\right)
\right|
_{\Lambda_{a}^{\ast}=0}\nonumber\\
& +\partial_{\theta}^{r}\varphi^{a}(\theta)\left\{  \frac{\partial
\mathsf{\Gamma}(\theta)}{\partial\varphi^{a}(\theta)}-\left(
\frac{\partial
X}{\partial\tilde{\varphi}{}^{a}(\theta)}\right)  \left(  {\varphi}^{b}%
+i\hbar(\mathsf{\Gamma}^{\prime\prime}{}^{-1})^{bc}\frac{\partial_{l}%
}{\partial\varphi^{c}},{i\hbar}\frac{\partial_{r}}{\partial(\partial_{\theta
}^{r}\varphi)}-\frac{\partial_{r}\mathsf{\Gamma}}{\partial(\partial_{\theta
}^{r}\varphi)}-\varphi^{\ast},\frac{\partial_{l}\mathsf{\Gamma}}%
{\partial\mathcal{I}}+\frac{\hbar}{i}\frac{\partial_{l}}{\partial\mathcal{I}%
},0\right)  \right\}  \nonumber\\
&  +\frac{1}{2}\left(
\mathsf{\Gamma}(\theta),\mathsf{\Gamma}(\theta)\right)
_{\theta}^{(\Gamma)}=0, \ \ \ \left[\mathsf{\Gamma}_{ab}^{\prime\prime}(\theta)\equiv\frac{\partial_{l}%
}{\partial\varphi^{a}(\theta)}\frac{\partial_{r}}{\partial\varphi^{b}(\theta
)}\mathsf{\Gamma}(\theta),\ \mathsf{\Gamma}_{ac}^{\prime\prime}%
(\theta)(\mathsf{\Gamma}^{\prime\prime}{}^{-1})^{cb}(\theta)=\delta_{a}\,^{b}\right]
,\label{51}%
\end{align}
which follow from the functional averaging of the
respective system for $W(\theta)$ and $X(\theta)$ in (\ref{42}),
as well as from integration by parts in the path integral, with
allowance for
$(\partial/\partial
\tilde{\varphi}^{\ast}\hspace{-0.1em}+\hspace{-0.1em}\partial/\partial{
\varphi}^{\ast})X(\theta)=0$.

For Abelian hypergauges $G_{A}\left(
(\Phi,\Phi^{\ast})(\theta)\right)
={\Phi}_{A}^{\ast}(\theta)-\partial\Psi(\Phi(\theta))/\partial\Phi^{A}%
(\theta)=0$ related to eq. (\ref{30}), with
$(\varphi,\varphi^{\ast},W)=(\Phi,\Phi^{\ast},S_{\mathrm{H};\mathrm{ext}})$,
$\partial_{\theta}^{r}{\Phi}{}^{A}=\mathcal{I}_{A}=0$, the
functional
$\mathsf{Z}(\partial_{\theta}\Phi^{\ast},\Phi^{\ast})(\theta)$ acquires the form%
\begin{equation}
\mathsf{Z}\left(  \partial_{\theta}\Phi^{\ast},\Phi^{\ast}\right)
(\theta)=\int d\Phi(\theta)\mathrm{exp}\left\{  \frac{i}{\hbar}\left[
{S}_{\mathrm{H}}^{\Psi}\left(  {\Gamma}(\theta),\hbar\right)  -((\partial
_{\theta}\Phi_{A}^{\ast})\Phi^{A})(\theta)\right]  \right\}  .\label{52}%
\end{equation}
For an HS with a Hamiltonian ${S}_{\mathrm{H}}^{\Psi}(\theta,\hbar)$ and
a solution $\check{\Gamma}(\theta)$ defined in $\Pi T^{\ast}\mathcal{M}_{\mathrm{ext}}$,
the BRST transformations are given by an anticanonical
(for a constant $\mu$) transformation:
\begin{equation}
\Gamma^{p}(\theta)\rightarrow\Gamma^{(1){}p}(\theta)=\exp\left[  \mu
s^{l({\Psi})}(\theta)\right]  \Gamma^{p}(\theta),\;s^{l({\Psi})}(\theta
)\equiv{\partial}/{\partial\theta}- \left(S_{\mathrm{H}}^{\Psi}%
(\theta,\hbar), \ \cdot \ \right)_{\theta}.\label{53}%
\end{equation}

From the permutation rule for the functional integral, $\varepsilon
(d\Phi(\theta))=0$,%
\[
\partial_{\theta}\int d\Phi(\theta)\mathcal{F}\left(  (\Phi,\Phi^{\ast
})(\theta),\theta\right)  =\int d\Phi(\theta)\left[
{\partial}/{\partial \theta}+(\partial_{\theta}V)(\theta)\right]
\mathcal{F}(\theta),\;\partial
_{\theta}V(\theta)=\left(\partial_{\theta}\Phi_{A}^{\ast}(\theta)\right){\partial
}/{\partial\Phi_{A}^{\ast}(\theta)}\,,
\]
with $i\hbar\partial_{\theta}^{r}\ln\mathsf{Z}(\theta)=(\partial
_{\theta}\Phi_{A}^{\ast}\partial_{\theta}^{r}\Phi^{A})(\theta)-\partial
_{\theta}^{r}\mathsf{\Gamma}(\theta)$, follow the relations
\begin{equation}
\left.  \partial_{\theta}\mathsf{Z}(\theta)\right|  _{\check{\Gamma}(\theta
)}=(\partial_{\theta}V)(\theta)\mathsf{Z}(\theta)=0,\;\left.  \partial
_{\theta}^{r}\mathsf{\Gamma}(\theta)\right|  _{\check{\Gamma}(\theta)}=\left(
\mathsf{\Gamma}(\Gamma(\theta)),\mathsf{\Gamma}(\Gamma(\theta))\right)
_{\theta}=0,\label{54}%
\end{equation}
which are implied by functional
averaging with respect to $\mathsf{Z}(\theta)$ and $\mathsf{\Gamma}(\theta)$,
\begin{equation}
\left.  \langle\partial_{\theta}^{r}\Gamma^{p}\rangle\right|  _{\mathsf{Z}%
}=\left(  \frac{\hbar}{i}\mathsf{Z}^{-1}\frac{\partial\mathsf{Z}(\theta
)}{\partial\Phi_{A}^{\ast}(\theta)},-\partial_{\theta}\Phi_{A}^{\ast}%
(\theta)\right)  ,\;\langle\partial_{\theta}^{r}\Gamma^{p}\rangle=\left(
\langle\Gamma^{p}(\theta)\rangle,\mathsf{\Gamma}(\langle\Gamma(\theta
)\rangle)\right)  _{\theta}=\partial_{\theta}^{r}\langle\Gamma^{p}%
\rangle,\label{55}%
\end{equation}
without the sign of averaging in (\ref{54}) for
$\check{\Gamma}^{p}(\theta)$ and ${\Gamma}^{p}(\theta)$. Formulae
(\ref{54}) relate the Ward identities in a theory with Abelian
hypergauges to the invariance of the generating functional of
Green's functions under superfield BRST transformations.

\section{Relation between Lagrangian Quantizations}

A relation between the conventional and $\theta$-local
quantizations can be established through the component form of
$\Gamma_{\mathrm{CL}}^{P},
\Gamma_{k}^{p_{k}}, \Lambda^{a}, \mathcal{I}_{a}, \Gamma_{k}^{p_{k}%
}(\theta)=\Gamma_{0k}^{p_{k}}+\Gamma_{1k}^{p_{k}}\theta$, $k=\mathrm{tot}$,
for $\theta=0$: $(\mathcal{M}$, $\mathcal{N}_{k}$, $\Lambda^{a}, \mathcal{I}%
_{a})\rightarrow(\widetilde{\mathcal{M}},\left.  \mathcal{N}_{k}\right|
_{\theta=0}=\{\Gamma_{0k}^{p_{k}}\},\lambda_{0}^{a},I_{0a})$. Besides the
condition $\theta=0$, a standard field model can be extracted
by eliminating the quantities $\partial_{\theta}\mathcal{A}^{I}(\theta)$, $\mathcal{A}%
_{I}^{\ast}(\theta)$ and the superfields $\mathcal{A}^{I}(\theta)$
with a wrong spin-statistics relation,
$\varepsilon_{P}(\mathcal{A}^{I})$ $\neq 0$. Such an
elimination can be realized\footnote{In theories of Yang--Mills type
a standard field model can also be extracted by special
\emph{horizontality conditions} \cite{Hull,NakanishiOjima}}
by eqs. (\ref{22}) and the conditions
$\mathrm{gh}(\mathcal{A}^{I})=-1-\mathrm{gh}(\mathcal{A}%
_{I}^{\ast})=0$, $(\varepsilon_{P})_{I}=0$.
For a restricted LSM of Section 3, a reduction to a model of the
multilevel formalism \cite{BatalinTyutin}
is achieved for vanishing
$\theta,\partial_{\theta}\varphi_{a}^{\ast
},\partial_{\theta}{\varphi}{}^{a},\varphi_{a}^{\ast},\mathcal{I}_{a}$.
Then the first-level functional integral $Z^{(1)}$ and its
symmetry transformations
\cite{BatalinTyutin}, with $\lambda_{0}^{a}$ instead\thinspace
of\thinspace $\pi^{a}$ \thinspace for\thinspace
Lagrangian\thinspace multipliers\thinspace of\thinspace
\cite{BatalinTyutin},
\begin{align*}
&  Z^{(1)}=\int d\lambda_{0}d\Gamma_{0}M(\Gamma_{0})\exp\left\{
\frac{i}{\hbar}\left(  W(\Gamma_{0})+G_{a}(\Gamma_{0})\lambda_{0}^{a}\right)
\right\}  ,\\
&  \left[  \delta\Gamma_{0}^{p},\delta\lambda_{0}^{a}\right]
=\left[  \left(
\Gamma_{0}^{p},-W+G_{a}\lambda_{0}^{a}\right)  ,-U_{cb}^{a}\lambda_{0}%
^{b}\lambda_{0}^{c}(-1)^{\varepsilon_{c}}+2i\hbar V_{b}^{a}\lambda_{0}%
^{b}+2(i\hbar)^{2}\tilde{G}^{a}\right]  \mu,
\end{align*}
coincide with $\left. \mathsf{Z}_{X}(0)\right|
_{\varphi_{0}^{\ast}=0}$ and its BRST transformations
generated by (\ref{44}) for
$R(\theta)=1$, under the identification $(\rho,\omega^{pq})(\Gamma
_{0})=(M,E^{pq})(\Gamma_{0})$ implying the coincidence of $\left.
{(\,\cdot\,,\,\cdot\,)_{\theta}}\right| _{\theta=0}$ and
$\Delta(0)$ with their counterparts of \cite{BatalinTyutin}.
The coincidence is implied by the choice of $X(\theta)$ as
\begin{equation}
X(\theta)=\left\{  G_{a}(\Gamma)\Lambda^{a}-\Lambda_{a}^{\ast}\left[
\frac{1}{2}U_{cb}^{a}(\Gamma)\Lambda^{b}\Lambda^{c}(-1)^{\varepsilon_{c}%
}-i\hbar V_{b}^{a}(\Gamma)\Lambda^{b}-(i\hbar)^{2}\tilde{G}^{a}(\Gamma
)\right]  \right\}  (\theta)+o(\Lambda^{\ast}),\label{57}%
\end{equation}
where $(V_{b}^{a},\tilde{G}^{a})(\theta)$ and
$(U_{cb}^{a},G_{a})(\theta)$ determine the unimodularity relations
\cite{BatalinTyutin}. A connection between the local quantization
and the generating functional of Green's function
$Z[J,\phi^{\ast}]$ of the BV method is evident after identifying
$\mathsf{Z}\left(  \partial_{\theta}\Phi^{\ast},\Phi^{\ast
}\right)  (0)=Z[J,\phi^{\ast}]$ in (\ref{52}),\thinspace
with\thinspace
the\thinspace action\thinspace${S}_{\mathrm{H}}^{\Psi}\left(  {\Gamma}%
_{0},\hbar\right)$ in (\ref{30}), (\ref{31}).

An arbitrary function $\mathcal{F}(\theta)$ $=$ $\mathcal{F}\left(
(\Gamma,\partial_{\theta}\Gamma)(\theta),\theta\right) \in
C^{\infty}\left(  \Pi T\mathcal{N}\times\{\theta\}\right)  $ can
be represented (in case $\Gamma^{p}=(\Phi^{A},\Phi_{A}^{\ast})$
see footnote 1) by a functional $F[\Gamma]$ of the superfield
scheme \cite{LMR},
\begin{equation}
F[\Gamma]=\partial_{\theta}\left[  \theta\mathcal{F}(\theta)\right]
=\mathcal{F}\left(  \Gamma(0),\partial_{\theta}\Gamma,0\right)  \equiv
\mathcal{F}(\Gamma_{0},\Gamma_{1})\,.\label{58}%
\end{equation}
This implies the independence of $F[\Gamma]$ from
$\partial_{\theta }^{r}\Gamma^{p}(\theta)$
in case $F(\theta)=F(\Gamma(\theta ),\theta)$. Formula (\ref{58})
allows one to establish a relation between the
objects $(\,\cdot\,,\,\cdot\,)_{\theta }^{\mathcal{N}}$ and
$\Delta^{\mathcal{N}}(\theta)$ in $C^{\infty }\left(
N\times\{\theta\}\right)$, with an extension to any $(\Gamma
,\omega^{pq},\rho)(\theta)$ of the flat operations
$(\,\cdot\,,\,\cdot\,)$, $\Delta$ in \cite{LMR}, which coincide
with their analogies of the BV method in case
$\Gamma^{p}=(\Phi^{A},\Phi_{A}^{\ast})$, $\omega^{pq}(\Gamma
(\theta))=\mathrm{antidiag}\left(
-\delta_{B}^{A},\delta_{B}^{A}\right)  $, $\rho(\theta)=1$, and in
the case of a different odd Poisson bivector,
$\tilde{\omega}^{pq}(\Gamma(\theta),\theta^{\prime})=(1+\theta^{\prime
}\partial_{\theta})\omega^{pq}(\theta)$. This correspondence is implied by%
\begin{align}
&  \left.  \left( {\mathcal{F}(\theta),\mathcal{G}(\theta)}\right)
{_{\theta}^{\mathcal{N}}}\right|  _{\theta=0}=\left(  F[\Gamma
],G[\Gamma]\right)  ^{\mathcal{N}}=\partial_{\theta}\left[
\frac{\delta_{r}F[\Gamma]}{\delta\Gamma^{p}(\theta)}\partial_{{\theta}%
^{\prime}}\left( \tilde{\omega}^{pq}(\Gamma(\theta),\theta^{\prime
})\frac{\delta_{l}G[\Gamma]}{\delta\Gamma^{q}(\theta^{\prime})}\right)
\right]  (-1)^{\varepsilon(\Gamma^{p})+1}, \label{59}\\
&  \left. {\Delta^{\mathcal{N}}(\theta)\mathcal{F}(\theta)}\right|
_{\theta=0}=\Delta ^{\mathcal{N}}F[\Gamma]
=\frac{1}{2}(-1)^{\varepsilon(\Gamma^{q})}%
\partial_{\theta}\partial_{\theta^{\prime}}\left[  \rho^{-1}[\Gamma
]\tilde{\omega}_{qp}(\theta^{\prime},\theta)\left(
\Gamma^{p}(\theta ),\rho\lbrack\Gamma]\left(
\Gamma^{q}(\theta^{\prime}), F[\Gamma]\right)
^{\mathcal{N}}\right)  ^{\mathcal{N}}\right]  ,\label{60}%
\end{align}
where $\left(  \rho\lbrack\Gamma],\tilde{\omega}_{pq}(\theta^{\prime}%
,\theta)\right)  =\left(  \rho(\Gamma_{0}),\theta^{\prime}\theta{\omega}%
_{pq}(\theta)\right)  $ and $\partial_{\theta}^{\prime\prime}\left[
\tilde{\omega}^{pd}(\theta^{\prime},\theta^{\prime\prime})\tilde{\omega}%
_{dq}(\theta^{\prime\prime},\theta)\right]
=\theta\delta^{p}{}_{q}.$ To establish the correspondence with
$(\,\cdot\,,\,\cdot\,)$ and $\Delta$ of \cite{LMR}
in (\ref{59}), (\ref{60}), one needs a relation between
superfield and component derivatives:
$\partial_{\theta }^{r}\Gamma^{p}(\theta)=(\lambda^{A},-(-1)^{\varepsilon_{A}}J_{A})$,
${\delta_{l}}/{\delta\Gamma^{p}(\theta)} =
(-1)^{\varepsilon(\Gamma^{p})}\left(
\theta{\delta_{l}}/{\delta\Gamma_{0}^{p}}-{\delta_{l}}/{\delta\Gamma_{1}^{p}%
}\right)$.

In general coordinates, the operators$\ \partial_{\theta}(V\pm U)^{\mathcal{N}%
}(0)$ in $\left.  \mathcal{N}=\Pi
T^{\ast}\mathcal{M}_{\mathrm{ext}}\right|
_{\theta=0}$, reduced to%
\[
\partial_{\theta}(V\pm U)(0)=\partial_{\theta}{\Phi}_{A}^{\ast}(\theta
)\partial/{\partial{\Phi}_{A}^{\ast}(0)}\pm\partial_{\theta}{\Phi}^{A}%
(\theta)\partial_{l}/{\partial{\Phi}^{A}(0)},
\]
coincide with the generalized sum and difference of $V$, $U$ in \cite{LMR},
for $\mathrm{t} = 1,2$:
\begin{align}
& \left.  \left.  \partial_{\theta}(V-(-1)^{\mathrm{t}}U)^{\mathcal{N}%
}(\theta)\mathcal{F}(\theta)\right|  _{\theta=0}=\left(  {\mathcal{S}%
^{\mathrm{t}}(\theta),\mathcal{F}(\theta)}\right)  {_{\theta}^{\mathcal{N}}%
}\right|  _{\theta=0} =(V-(-1)^{\mathrm{t}}U)^{\mathcal{N}}F[\Gamma]=\left(  S^{\mathrm{t}%
}[\Gamma],F[\Gamma]\right)  ^{\mathcal{N}},\nonumber\\
& \mathcal{S}^{\mathrm{t}}(\theta)=(\partial_{\theta}\Gamma^{p})\omega
_{pq}^{\mathrm{t}}(\Gamma(\theta))\Gamma^{q}(\theta),\;S^{\mathrm{t}}%
[\Gamma]=\partial_{\theta}\left\{  \Gamma^{p}(\theta)\partial_{\theta^{\prime
}}\partial_{\theta}\left[  \tilde{\omega}_{pq}^{\mathrm{t}}(\theta
,\theta^{\prime})\Gamma^{q}(\theta^{\prime})\right]  \right\}  =\mathcal{S}%
^{\mathrm{t}}(0),\label{61}%
\end{align}
where the $\vec{\varepsilon}$-bosonic quantities
$\mathcal{S}^{\mathrm{t}}(\theta)$ or $S^{\mathrm{t}}[\Gamma]$ must
obey certain differential equations providing the
anticommutativity of the operators $\{\Delta^{\mathcal{N}},\partial _{\theta }V^{\mathcal{N}}%
,\partial _{\theta }U^{\mathcal{N}}\}(\theta)$, while
$\omega_{pq}^{\mathrm{t}}(\theta)$ and $\tilde{\omega}_{pq}^{\mathrm{t}%
}(\theta,\theta^{\prime})$, identical to $\omega_{pq}(\theta)$ and
$\tilde{\omega}_{pq}(\theta,\theta^{\prime})$ in case
$\mathrm{t}=1$, are given by%
\[
\tilde{\omega}_{pq}^{\mathrm{t}}(\theta,\theta^{\prime})=\theta\theta^{\prime
}\omega_{pq}^{\mathrm{t}}(\theta^{\prime})=-(-1)^{\mathrm{t}+\varepsilon
(\Gamma^{p})\varepsilon(\Gamma^{q})}\tilde{\omega}_{qp}^{\mathrm{t}}%
(\theta^{\prime},\theta),\ \omega_{pq}^{\mathrm{t}}(\theta)=(-1)^{\varepsilon
(\Gamma^{p})\varepsilon(\Gamma^{q})+\mathrm{t}}\omega_{qp}^{\mathrm{t}}%
(\theta)\,.
\]

\section{Summary}

We have proposed a $\theta$-local description of an arbitrary
reducible superfield theory as a natural extension of a standard
gauge theory with classical fields $A^{i}$ to a superfield
model defined on extended cotangent $\left\{
\mathcal{A}^{I},\mathcal{A}_{I}^{\ast}\right\}  (\theta)$ and
tangent $\left\{
\mathcal{A}^{I},\partial_{\theta}\mathcal{A}^{I}\right\} (\theta)$
odd bundles in respective Hamiltonian and Lagrangian
formulations. It is shown that the conservation, under the
$\theta$-evolution, of a Hamiltonian action $S_{H}\left(
(\mathcal{A},\mathcal{A}^{\ast })(\theta),\theta\right)  $, or,
equivalently, of an odd analogue of the energy, $S_{E}\left(
(\mathcal{A},\partial_{\theta}\mathcal{A})(\theta ),\theta\right)
$, is equivalent to a respective Hamiltonian or Lagrangian master
equation. We have proposed a $\theta$-local
description of Lagrangian quantization in non-Abelian hypergauges
for a reducible gauge model extracted from a general superfield
model by conditions of the $\theta$-independence of the classical
action and the vanishing of ghost number for
$\mathcal{A}^{I}(\theta)$ and the action.
To investigate the BRST
invariance and gauge-independence of the generating functionals of
Green's functions, we have used \emph{two equivalent}
Hamiltonian-like systems, defined in terms of a $\theta$-local
antibracket and an even Poisson bracket, respectively.
These systems permit a simultaneous description of BRST
transformations and continuous (anti)canonical-like
transformations.
We have established the coincidence of the
first-level functional integral $Z^{(1)}$ in the first-level formalism
\cite{BatalinTyutin}
with the local vacuum function of the suggested quantization
scheme, as well as the coincidence of the generating functional
$Z(\phi^{\ast},J)$ of the BV method with
$\mathsf{Z}(\partial\Phi^{\ast},\Phi^{\ast})(0)$.

\textbf{Acknowledgments} The authors are grateful to P.M. Lavrov for useful
discussions. D.M.G. thanks the foundations FAPESP and CNPq for permanent
support. P.Yu.M. is grateful to FAPESP.

\end{document}